\documentclass[fleqn,usenatbib,useAMS]{mnras}

\usepackage{graphicx}	
\usepackage{amsmath}	
\usepackage{amssymb}	
\usepackage{multicol}        
\usepackage{bm}		
\usepackage{pdflscape}	
\usepackage{comment}
\usepackage{soul}

\usepackage[T1]{fontenc}
\usepackage{ae,aecompl}

\usepackage{newtxtext,newtxmath}

\usepackage{scalerel}
\usepackage{tikz}
\usetikzlibrary{svg.path}
\definecolor{orcidlogocol}{HTML}{A6CE39}
\tikzset{
	orcidlogo/.pic={
		\fill[orcidlogocol] svg{M256,128c0,70.7-57.3,128-128,128C57.3,256,0,198.7,0,128C0,57.3,57.3,0,128,0C198.7,0,256,57.3,256,128z};
		\fill[white] svg{M86.3,186.2H70.9V79.1h15.4v48.4V186.2z}
		svg{M108.9,79.1h41.6c39.6,0,57,28.3,57,53.6c0,27.5-21.5,53.6-56.8,53.6h-41.8V79.1z M124.3,172.4h24.5c34.9,0,42.9-26.5,42.9-39.7c0-21.5-13.7-39.7-43.7-39.7h-23.7V172.4z}
		svg{M88.7,56.8c0,5.5-4.5,10.1-10.1,10.1c-5.6,0-10.1-4.6-10.1-10.1c0-5.6,4.5-10.1,10.1-10.1C84.2,46.7,88.7,51.3,88.7,56.8z};
	}
}

\newcommand\orcidicon[1]{\href{https://orcid.org/#1}{\mbox{\scalerel*{
				\begin{tikzpicture}[yscale=-1,transform shape]
					\pic{orcidlogo};
				\end{tikzpicture}
			}{|}}}}

\DeclareRobustCommand{\VAN}[3]{#2}
\let\VANthebibliography\thebibliography
\def\thebibliography{\DeclareRobustCommand{\VAN}[3]{##3}\VANthebibliography}

\title[Scale of homogeneity using multi-fractal approach]{Investigating Cosmic Homogeneity Using Multi-fractal Analysis of the SDSS-IV eBOSS DR16 Quasar Catalog}

\author[Goyal et al.]{
Priya Goyal,$^{1}$\thanks{E-mail:priyagoyal@kias.re.kr}
Sunil Malik
\orcidicon{0000-0003-4147-626X}$^{2,3}$
\thanks{Corresponding Author: sunil.malik@uni-potsdam.de}
Jaswant K. Yadav$^{4}$,
T.R. Seshadri$^{5}$
\\
\\
$^{1}$Korea Institute for Advanced Study, 85-Hoegiro, Dongdaemun-gu, Seoul-02455, South korea\\
$^{2}$Institute fur Physik und Astronomie Universitat Potsdam, Golm Haus 28, D-14476 Potsdam, Germany\\
$^{3}$Deutsches Elektronen-Synchrotron DESY, Platanenallee 6, 15738 Zeuthen, Germany,\\
$^{4}$Department of Physics and Astrophysics, Central University of Haryana, Mahendergarh-123031 India\\
$^{5}$ Department of Physics and Astrophysics, University of Delhi, Delhi-110007, India.
}

\date{Accepted XXX. Received YYY; in original form ZZZ}

\pubyear{2024}

\begin{document}
\label{firstpage}
\pagerange{\pageref{firstpage}--\pageref{lastpage}}
\maketitle

\begin{abstract}
 We analyze the volume-limited subsamples extracted from the sixteenth data release of the SDSS-IV eBOSS quasar survey spanning a redshift interval of $0.8 < z < 2.2$, to estimate the scale of transition to homogeneity in the Universe. The multi-fractal analysis used for this purpose considers the scaling behavior of different moments of quasar distribution 
 in different density environments. This analysis gives the spectrum of generalized dimension $D_q$, where positive values of $q$ characterize the scaling behavior in over-dense regions and the negative ones in under-dense regions. We expect fractal correlation dimension $D_q(r) = 3$, for a homogeneous, random point distribution in 3-Dimensions. The fractal correlation dimension $D_q(r)$, corresponding to $q=2$ obtained in our study stabilizes in the range (2.8-2.9) for scales $r>80$ $h^{-1}$ Mpc. The observed quasar distribution shows consistency with the simulated mock data and the random distribution of quasars within one sigma. Further, the generalized dimension spectrum $D_q(r)$ also reveals transition to homogeneity beyond $>110$ $h^{-1}$ Mpc, and the dominance of clustering at small scales $r<80$ $h^{-1}$ Mpc. Consequently, our study provides strong evidence for the homogeneity in SDSS quasar distribution, offering insights into large-scale structure properties and, thus can play a pivotal role in scrutinizing the clustering properties of quasars and its evolution in various upcoming surveys such as Dark Energy Spectroscopic Instrument (DESI) and Extremely Large Telescope (ELT).  

\end{abstract}

\begin{keywords}
cosmology: large-scale structure of Universe - Homogeneity scale – methods: statistical- Multifractal analysis – galaxies: quasars: general.
\end{keywords}



\section{Introduction}
Galaxy redshift surveys are the most direct methods for exploring the distribution of the Universe's large-scale structure (LSS) across space and time. These surveys tell us in the most straightforward way what our Universe looks like. It is now confirmed that the Universe contains structures of varying size and shapes, from galaxy to galaxy clusters, filaments, voids, walls, collectively called the cosmic web~\citep{Bond1996}. These structures have formed and evolved from tiny density perturbations in the very early Universe by hierarchical growth driven primarily by gravity \citep{2008arXiv0804.2258L,vandeWeygaert2009}. Despite the existence of clustering at small scales, we expect the Universe to be homogeneous on large scales \citep{yadav2005, Hogg:2004vw,Scrimgeour:2012wt}.
One of the foundational principles underlying the standard theory of cosmology, known as the $\Lambda$CDM model (Cold Dark Matter with a cosmological constant, $\Lambda$), is the "Cosmological Principle". This principle posits that our Universe exhibits statistical homogeneity and isotropy on large scales (typically $> 100 h ^{-1}$ Mpc ) \citep{1937RSPSA.158..324M,Clarkson1999,Katanaev2015MPLA,Camacho2022JCAP}. 
Testing these assumptions in the era of the vast current and upcoming observational data is an essential robustness check of the standard cosmological model. The assumption of isotropy has been tested using various types of data sets like distribution of radio source counts~\citep{2019JCAP...09..025B}, X-ray surveys of galaxy clusters~\citep{2018A&A...617C...2V,2020A&A...636A..15M}, the temperature and polarization anisotropies of the Cosmic Microwave Background Radiation (CMBR)~\citep{refId0}, by employing different statistical methods. These high-precision experiments support the statistical isotropy of the Universe. However, certain observations have hinted at anomalies challenging the assumption of isotropy~\citep{2023CQGra..40i4001K}. For instance, anomalies like hemispherical asymmetry and point parity symmetry violation
have been found in the Planck data~\citep{refId0}, along with statistically
significant higher dipole amplitude measured in the distribution of
quasars compared to the CMBR value~\citep{Secrest_2021}.
Although numerous tests of the isotropy hypothesis have been performed in a direct manner through observations, the assumption of cosmic homogeneity can only be indirectly tested, this is because we observe down the past light cone and not on the time-constant hyper-surfaces~\citep{2010CQGra..27l4008C,2012CRPhy..13..682C}. Thus, investigating the statistical homogeneity of the Universe on very large scales is a challenging endeavour \citep{2011RSPTA.369.5115M}. 

The methodology for investigating the homogeneity involves analyzing the clustering properties of galaxies, quasars, and other tracers of LSS across different length scales. Currently, most of the analyses performed found a transition scale from a locally clustered to a smooth, statistically homogeneous Universe, using galaxy or quasar counts, in the interval $70 < r_{h} < 150$ $h^{-1}$ Mpc~\citep{Hogg:2004vw,yadav2005, 10.1111/j.1745-3933.2009.00738.x,Scrimgeour:2012wt,10.1093/mnras/stv1994,Laurent:2016eqo,10.1093/mnrasl/slw145,Goncalves:2018sxa, Goncalves_2021,Pandey_2021}.
However, some studies still argue against the existence of any such homogeneity scale \citep{Labini:2009ke, Labini:2011dv, Park:2016xfp}. Any deviations from the assumptions of isotropy and homogeneity can have significant implications for our current understanding of the Universe \citep{2004LRR.....7....8L}. Therefore, it is crucial to rigorously test these fundamental assumptions underlying our standard cosmological model using a variety of existing and upcoming cosmological data, utilizing the rich statistical tools available today~\citep{2015A&A...582A.111L,2019JCAP...05..048J,2022JCAP...04..044C}.

At low redshifts, $z <1$, galaxies are used as direct tracers of the matter density field, while at high redshifts $z > 2$, clouds of neutral hydrogen in the Lyman$-\alpha$ Forest, as illuminated by background quasar-light, are similarly used to test the assumptions of homogeneity and isotropy of the Universe~\citep{2019MNRAS.489.3966Z}. Quasar distribution provides a reliable tracer population in an intermediate redshift range~\citep{refud0}, since they are the brightest objects, hence can be observed more easily at cosmological distances. While it is true that quasars are biased tracers of matter density field, this drawback is offset by the advantage that quasars sample a very large volume of space and so are suited for testing cosmic homogeneity over large scales. Traditional methods like $2-$point correlation function ~\citep{2017JCAP...07..017L} and power spectrum approach ~\citep{2018MNRAS.477.1604G,Hou:2020rse, Zhao:2021ahg} have been employed to study the clustering of quasar distribution in the Sloan Digital Sky Survey (SDSS)~\citep{2020MNRAS.498.3470W,2022MNRAS.511.5492Z}. These methods assume the distribution of quasars to be homogeneous within the survey boundaries and therefore, can not be used to test the principle of homogeneity of the quasar distribution.
Most statistical methods used to measure homogeneity rely on the "counts-in-spheres" technique. This involves counting the number of sources (galaxies/quasars) within spheres of certain radius, centred on each galaxy/quasar. By averaging these counts over a large number of centers, the expectation is that, for a random distribution of points, the average number count scales proportionally to the volume of the sphere~\citep{1995PhR...251....1B}. This formalism forms the basis of fractal analysis where the scaling exponent of the average number count is proportional to the fractal Dimension~\citep{mandelbrot1982fractal}. The most commonly used one is the correlation dimension, $D_2(r)$, which quantifies the scaling of the two-point correlation function. The counts in sphere is closely related to the volume integral of the two-point correlation function, $\xi(r)$. 
In a situation where $\xi(r)$ obeys a power-law behaviour, i.e. $\xi(r ) = {(r /{r_0} )}^{\gamma}$ , the correlation dimension is given by, $D_2 =3-\gamma$, on scales $r < {r_0}$, where 3 denotes the ambient dimension of the space. Here, the ambient dimension by definition is the dimension of space in which the particle distribution is embedded, for example, it will be 3, for a 3-D spatial distribution of quasars or galaxies~\citep{1999A&A...351..405B}. Several previous studies~\citep{Scrimgeour:2012wt,10.1093/mnras/stv1994,Laurent:2016eqo,10.1093/mnrasl/slw145, Goncalves:2018sxa, Goncalves_2021} have employed fractal correlation dimension analysis on the SDSS galaxy and quasar distributions to infer the cosmic homogeneity scale. They conclude that the large-scale homogeneity assumption is consistent with the largest spatial distribution of quasars currently available. Primordial non-Gaussianity~\citep{2015A&A...582A.111L} and non-linear evolution at later stages of evolution of the matter density field leads to the non-Gaussian distribution of tracers of the density field. Therefore, a distribution of quasars in the Universe is not necessarily a Gaussian distribution and hence may not be completely specified by its $2-$point correlation function (or equivalently $D_2$) and thus we must investigate the higher-order correlations.

In this study, we employ the concept of multi-fractal dimension also known as Minkowski-Bouligand dimension ($D_q$)~\citep{HENTSCHEL1983435,1999A&A...351..405B} on the spatial distribution of quasars. Here $q$ is an integer that can take positive and negative values. These dimensions quantify the scaling behavior of different moments of the counts-in-spheres in the distribution of points and are related to a combination of n-point correlation functions~\citep{1995PhR...251....1B}. A monofractal is a special case of multi-fractal where the value of $D_q$ is independent of $q$. The multi-fractal dimension of matter distribution can be used to test the hypothesis of homogeneity. In practice, for any given tracer distribution, one computes the multi-fractal dimension of the point distribution, and the length scale above which the multi-fractal dimension is equal to the ambient dimension of the space can be considered as the scale of homogeneity of that distribution. 

The focus of this work is to investigate the homogeneity scale in the distribution of quasars. We perform the multi-fractal analysis to study the scaling properties (counts in the sphere) of the quasar distribution in the latest SDSS-IV DR16 eBOSS survey~\citep{2020MNRAS.498.2354R} and test if it is consistent with homogeneity on large scales. 
This data set covers a redshift interval, $0.8 < z < 2.2$, thus providing significant volume in the Universe to investigate the clustering properties. We divide the sample into four redshift bins spanning the whole redshift coverage of the survey and use counts in the sphere and its different moments to study if there is evidence for the existence of such a homogeneity scale. The generalized spectrum of fractal dimension ($D_q$) obtained from multi-fractal analysis complements the more commonly used correlation dimension ($D_2$) to extricate the underlying homogeneity scale from the distribution of quasars. 

This paper is organized as follows. In Section~\ref{data}, we briefly describe the observational data used in our analysis. In Section~\ref{sec:maths}, we describe our methodology. We present our findings in Section~\ref{analysis} and 
the Section~\ref{discuss} gives a brief discussion of the results. Finally, the main conclusions of our study are summarised in Section~\ref{conclusion}.

\section{Observational Data}
\label{data}
We have utilized the SDSS-IV quasar catalog available in the 16th Data Release (DR16) of the extended Baryon Oscillation Spectroscopic Survey (eBOSS) survey.\footnote{The SDSS-IV DR16 eBOSS clustering catalogs (NGC $\&$ SGC) used in this analysis are available at the following URL: \url{https://data.sdss.org/sas/dr16/eboss/lss/catalogs/DR16/}} \cite{2020MNRAS.498.2354R}. It consists of about $218,209$ quasars located in the Northern Galactic Cap (NGC) region and $125,99$ quasars in the Southern Galactic Cap (SGC) region. These quasars cover a redshift range of $0.80 < z < 2.20$, representing a substantial depth of observations. The observed quasar distribution spans an effective area of approximately $4699 \ \text{deg}^2$. It is worth noting that this survey has a mean completeness of $C_{\text{comp}} \sim 0.98$ for both regions. 

To investigate the scale of homogeneity across different epochs, we divided the sample into four distinct redshift bins. The mean redshift and the corresponding number of sources in each bin are presented in Table~\ref{tab:example_table}. We have applied a threshold on the minimum number of quasars in the subsamples. This threshold is determined based on the quasar subsample in the highest redshift bin. The width of each redshift bin is determined to be the broadest possible range encompassing an almost equal number of data points while ensuring no overlap between adjacent redshift bins. The number of quasars is kept similar across all the redshift bins so that the Poisson fluctuation which is proportional to $1/\sqrt{N}$ is similar across all bins. This is similar to what was followed in previous works~\citep{Goncalves:2018sxa,Goncalves_2021}.

In order to have an estimate of the uncertainties due to various observational systematic effects, we have utilized the Extended Zel’dovich (EZ) mock catalogs~\citep{2021MNRAS.503.1149Z}, which has the same clustering property of the eBOSS DR16 quasar sample. The EZmock algorithm uses Zel’dovich approximation to construct the density field at a given redshift, and populate matter tracers (haloes/galaxies/quasars) in the field with a parameterized modelling of tracer bias. This effective bias description includes linear, nonlinear, deterministic, and stochastic effects, which have to be calibrated with clustering statistics from observations or N-body simulations, including typically the two-point correlation function (2PCF), power spectrum, and bispectrum. Moreover, various geometrical survey features are applied to the mocks, including survey footprints, veto masks, and radial distributions. In addition, photometric and spectroscopic systematic effects associated with the observations are migrated to the EZmock catalogs to have robust estimates of the covariance matrices for Baryonic Acoustic Oscillation (BAO) and Redshift Space Distortion (RSD) analysis. These EZmock catalogs display good agreements with the observational data regarding various clustering statistics, especially at small scales \citep{2021MNRAS.503.1149Z}. 
We conducted an analysis on approximately 40 mock realizations using the same methodology employed for the observed quasar data.

In addition, for quasar observation sample in each region (NGC and SGC), a corresponding random sample~\citep{2020MNRAS.498.2354R} is generated that is at least 40 times as dense and approximates the respective 3D (RA, DEC, redshift) selection functions of the observed distribution. Various weights are provided for both the data and random samples to ensure that the latter matches the selection function and optimizes the signal-to-noise of the clustering measurements. More details about creating LSS clustering catalogs for the eBOSS DR16 quasar samples and the random catalogs can be found in  \cite{2020MNRAS.498.2354R}. The analysis of random samples
(which are Poisson sampled homogeneous distribution by construction but with statistical fluctuations) is important to check the scales at which
real distribution is close enough to the random ones. We utilized 40 samples of these random catalogs in our analysis. In the following section, we give a brief description of the multi-fractal analysis method.
\begin{table}
	\centering
	\caption{The table shows the redshift interval and the respective number of quasars in each bin for both NGC and SGC regions.}
	\label{tab:example_table}
	\begin{tabular}{|l|c|c|c|} 
	    \hline
	    z interval & $\bar{z}$ & $N_q$ (NGC)& $N_q$ (SGC)\\
		\hline
		0.800-1.135 & 0.967  & 42488 & 23791 \\
		1.225-1.460 & 1.342 & 42720 & 24327 \\
		1.560-1.800  & 1.680  & 42771 & 25043\\
		1.890-2.200  & 2.045 & 42574 & 25131\\
		\hline
	\end{tabular}
 \label{tab:list}
\end{table}

\section{Methodology}
\label{sec:maths} 
Past analysis of galaxy redshift surveys suggest that on small scales matter distribution in the Universe resembles a fractal distribution, characterised by a parameter called fractal dimension~\citep{WENZHENG1988269,PEEBLES1989273,Teles2022}. There are different ways to calculate fractal dimension, and the Minkowski-Bouligand dimension is of particular relevance in the context of cosmology for analyzing distributions of galaxies or quasars \citep{MS}. We now give an overview of the method adopted to measure counts in the sphere in the 3-D distribution of quasars.

\begin{figure*}
\includegraphics[width=\columnwidth]{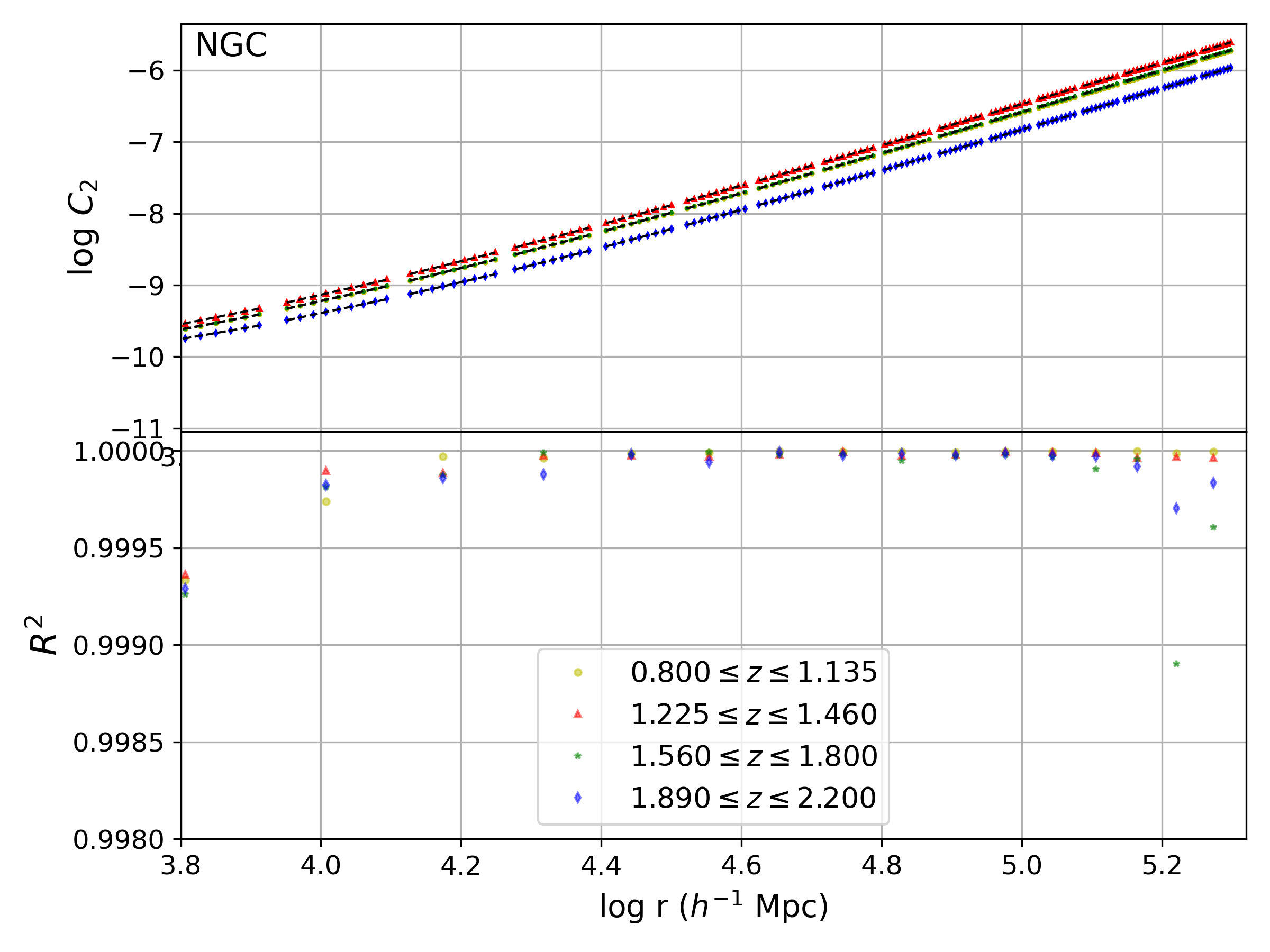}
    \includegraphics[width=\columnwidth]{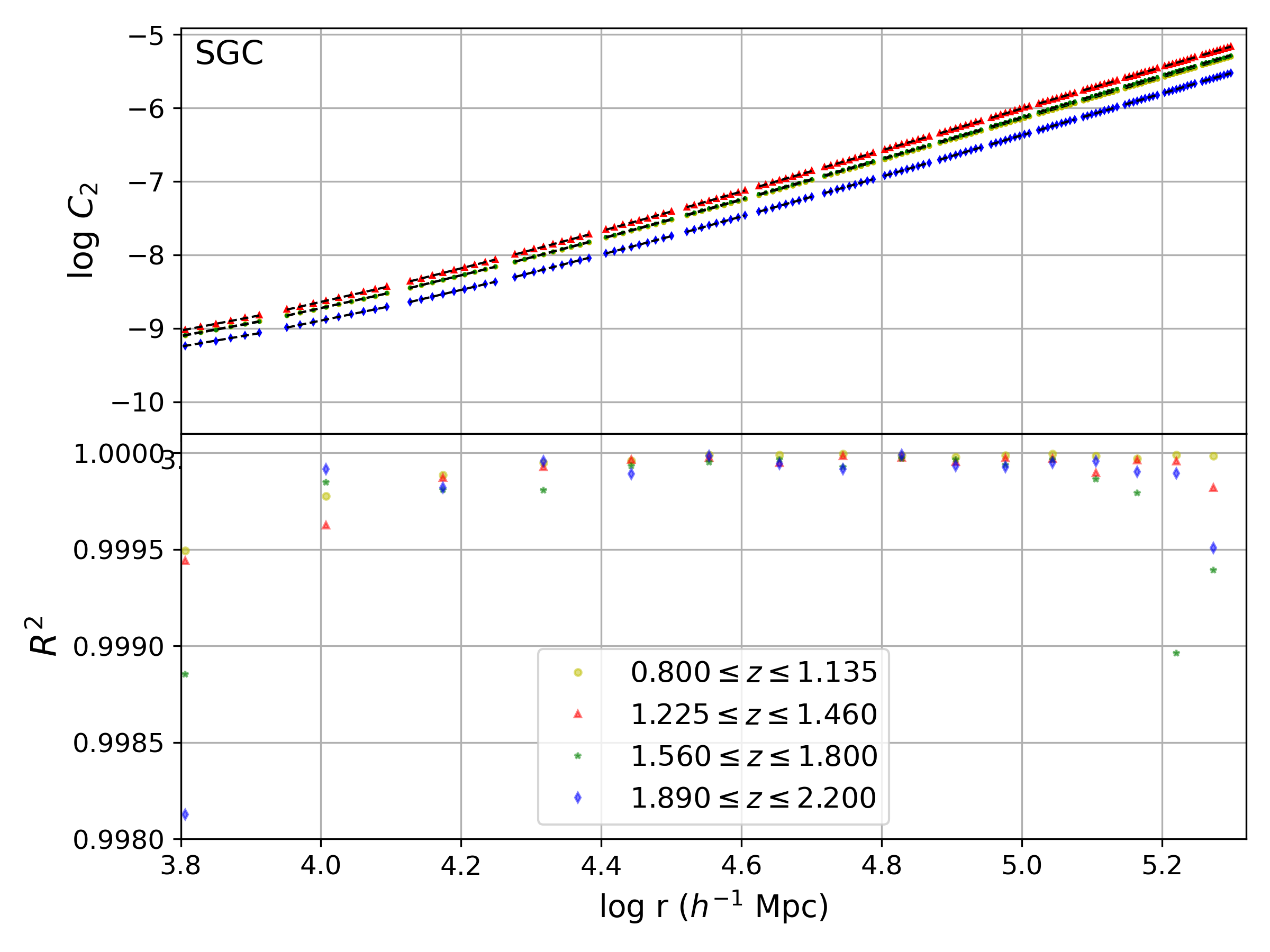}
    \caption{This figure shows log $C_q (r)$ versus log r for q=2, for the DR16 eBOSS quasar data in both the NGC (upper left panel) and SGC region (upper right panel). Each curve represents one of the four non-continuous redshift bins considered in our analysis, characterized by mean redshifts of $\bar{z}=0.967$ (yellow circles), $\bar{z}=1.342$ (red triangles), $\bar{z}=1.680$ (green stars), and $\bar{z}=2.045$ (blue diamonds). Beneath the data points, we display the best fit linear model in black, extending across various length scales, which implies the power law scaling behaviour of $C_2$(r) with a positive exponent. Furthermore, in the lower panels of each plot, we present the goodness-of-fit parameter ($R^2$) as a function of scale. This metric quantifies the robustness of our linear fitting. Notably, its value exceeds 0.998 across all length scales, indicating that the assumption of power-law scaling behavior of $C_2$ is well-justified.}
    \label{fig:C2_NGC}
\end{figure*}

If we denote by ${\vec {x_i}}$ and ${\vec {x_j}}$, the position vectors of the $i$th and $j$th quasar, the distance between them is given by,
\begin{equation}
{\mid{\vec {x_i}}-{\vec {x_j}\mid}} = \sqrt{d(z_{i})^{2} + d(z_{j})^{2} -2d(z_{i}) d(z_{j}) cos(\theta_{ij})}
\end{equation}
where $cos(\theta_{ij}) = sin(\delta_{i}) sin(\delta_{j}) + cos(\delta_{i})cos(\delta_{j}) cos(\alpha_{i} - \alpha_{j})$,  and $\alpha$, $\delta$, and $d(z)$ are Right Ascension (RA), Declination (DEC), the
radial comoving distance of the source, respectively. 

The $d(z)$ is given by,
\begin{equation}
    d(z) = \int_{0}^{z} \frac{c dz^{'}}{H(z^{'})}
\end{equation}
and can be computed using the fiducial cosmological parameters fixed at $\Omega_{m0} = 0.30712$,\, $\Omega_{b0} = 0.048252$, $\Omega_{\Lambda 0} = 0.644628$ and
$H_0 = 67.77$ km {\rm s$^{-1}$ Mpc$^{-1}$}~\citep{2014A&A...571A..16P}. Therefore, the number of quasars within a sphere of the comoving radius $r$ centered at the ith quasar is given by~\citep{yadav2005}, 
\begin{equation}
    n_{i}(r) = \sum_{j=1}^{N} \Theta(r - |x_{i}- x_{j}|)
\end{equation}
where $\Theta(x)$ represents the the Heaviside function, which is defined as $\Theta(x)$ = 0 for $x < 0$ and $\Theta(x)$ = 1 for $x > 0$. Furthermore, averaging $n_i(r)$ by taking M number of different quasars (inside the considered redshift bin  after making edge correction) as centers and dividing by the total number of
quasars defines the correlation integral as;
\begin{equation}
    C_{2}(r) = \frac{1}{MN}\sum_{i=1}^{M} n_{i}(r)
    \label{eqn:counts}
\end{equation}
where N and M are the total numbers of quasars and the number of possible quasar centres at each $r$, respectively. For all considered values of r, $M<N$ (due to quasars removed from the edges of the redshift bin). As the radius of the sphere increases, the number of quasars available as centers decreases for larger spheres, i.e. the value of M decreases as we increase r.
Here, $C_{2}(r)$ can be understood as the probability of finding a quasar within a sphere of radius $r$ centered on another quasar. If $C_2 (r)$ exhibits a power-law scaling relation\footnote{Since fractals are self-similar structures/distributions, a fractal analysis is possible only if $C_2$ (Or for that matter all $C_q$) follow a power law scaling. However, throughout the entire observed range it may not be so. This is because, at small scales, the discreteness of the distribution shows up significantly and, at very large scales, the finiteness of the volume of sample space is a constraint. Further, the evolution of $C_2$ with time can lead to a broken power-law. Hence, we identified power-law regions and applied our technique there as these are the faithful representation of the underlying distributions devoid of the above artifacts.}, $C_{2}(r) \propto r^{D_{2}}$, the exponent $D_2$ is defined to be the correlation dimension. 

\begin{equation}
    D_{2}(r) = \frac{d \log C_{2}(r)}{d \log r}
    \label{eqn:Dimension}
\end{equation}

The scaling behaviour of $C_2$ can be different at different scales, therefore, we expect the correlation dimension $D_2$ to be a function of scale. For the special case of a homogeneous distribution, we see that the correlation dimension equals the ambient dimension, i.e. $D_2 \sim 3$.

The full statistical quantification of a fractal distribution requires a hierarchy of scaling indices similar to higher-order correlations required to characterize all the statistical properties of the large-scale structure distribution~\citep{10.1093/mnras/stw3234,Marin_2011,2011ApJ...726...13M}. It is in this context that the multi-fractal analysis used here is important. It provides a spectrum of generalized dimension $D_q$, the Minkowski–Bouligand dimension. These dimensions quantify the scaling behavior of different moments of the counts-in-spheres and are related to a combination of n-point correlation functions~\citep{1995PhR...251....1B}. We generalize Equation~\ref{eqn:counts} to 
\begin{equation}
    C_{q}(r) = \frac{1}{MN}\sum_{i=1}^{M} [n_{i}(r)]^{q-1},
    \label{eqn:multi}
\end{equation}
which is then used to define the Minkowski–Bouligand dimension $D_q$ as,
\begin{equation}
    D_{q}(r) = \frac{1}{q-1} \frac{d \log C_{q} (r)}{d \log r}
    \label{eqn:muti_dimension}
\end{equation}
Here $q$ is an integer that can take positive and negative values. The $C_q(r)$ for a fixed $r$ will be dominated by the contribution from high (low) density regions if we consider high positive (negative) values of $q$. The varying values of $q$ in our analysis also help us determine the scaling behavior of different moments of counts in spheres thereby giving a combined view of the scale of homogeneity of the universe. We have considered q in the range, $-5\leq q \leq 5$ in our study. The finite number of quasars restricts us from considering any arbitrary large values of $|q|$.
The values of $D_q$ at $q=1$ and $q=2$ correspond to the box-counting dimension and correlation dimension, respectively. The case of $q=1$ has to be dealt with care by taking a suitable limiting case of $q$ tending to $1$. For a multi-fractal distribution, the values of $D_q$ will be different for different values of q. However, for a mono-fractal, $D_q$ is constant, independent of $q$. And for a homogeneous distribution in 3-Dimension, the value of $D_q$ should be equal to 3 for all q. We now apply this multi-fractal methodology to the SDSS-IV DR16 eBOSS quasar distribution, EZmocks and the random quasar distribution to investigate the cosmic homogeneity scale in the quasar distribution. The random samples are homogeneous by construction. So, to determine the scale of transition to homogeneity, we compare the results for observed quasar distribution with that from the random sample. 

\section{Analysis and Results}
\label{analysis}
\subsection{Behaviour of Correlation Integral and Correlation Dimension}
To estimate the homogeneity scale within each of the four redshift bins outlined in Table~\ref{tab:list} for both NGC and SGC observed quasar distributions, we first compute the correlation integral $C_2(r)$ as defined in Equation~\ref{eqn:counts}.
In the top left (top right) panel of Figure~\ref{fig:C2_NGC}, we show the variation of the logarithm of $C_2$ as a function of the logarithm of $r$ in the r range $(20-140)$ $h^{-1}$ Mpc for the four redshift bins in the NGC (SGC) region. These plots illustrate the relationship between log$C_2$ and log $r$ which appears to be linear. It can therefore be asserted that $C_2$ increases monotonically as a power law in $r$. However, the exponent of the power law in $r$ varies across the range of $r$ investigated in our study. To deduce the power law exponent, we have performed the linear fitting of log $C_2$ versus log $r$ across intervals of 10 Mpc range within the full range of [20-140 $h^{-1} \ Mpc$], where the slopes will represent the correlation dimension $D_2$ value as defined in Equation~\ref{eqn:Dimension}. Furthermore, to assess the robustness of the linear fitting, we computed $R^{2}$-squared correlation coefficient, which serves as one of the indicators of goodness of fit for linear regression models~\citep{r2_paper}. 
The $R^2$ provides a useful measure of how well a model fits the data, in terms of (squared) distance from points to the best-fitting line.  Its value ranges from 0 to 1, where values closer to 1 indicate a better fit of the model to the data. We plot the $R^2$ coefficient for each 10 Mpc interval in r for all the four redshift bins in the bottom panels of Figure~\ref{fig:C2_NGC}. We note that for the complete $r$ range, its value exceeds 0.998, suggesting that the relationship between log $C_2$ versus log $r$ is well approximated by a linear model or $C_2$ exhibits a power law behaviour in r. 
Additionally, we computed $C_2{(r)}$ for 40 realizations of EZmocks and 40 corresponding random realizations in a similar manner as done for the observed sample.

\begin{figure*}
    \includegraphics[width=0.89\columnwidth]   {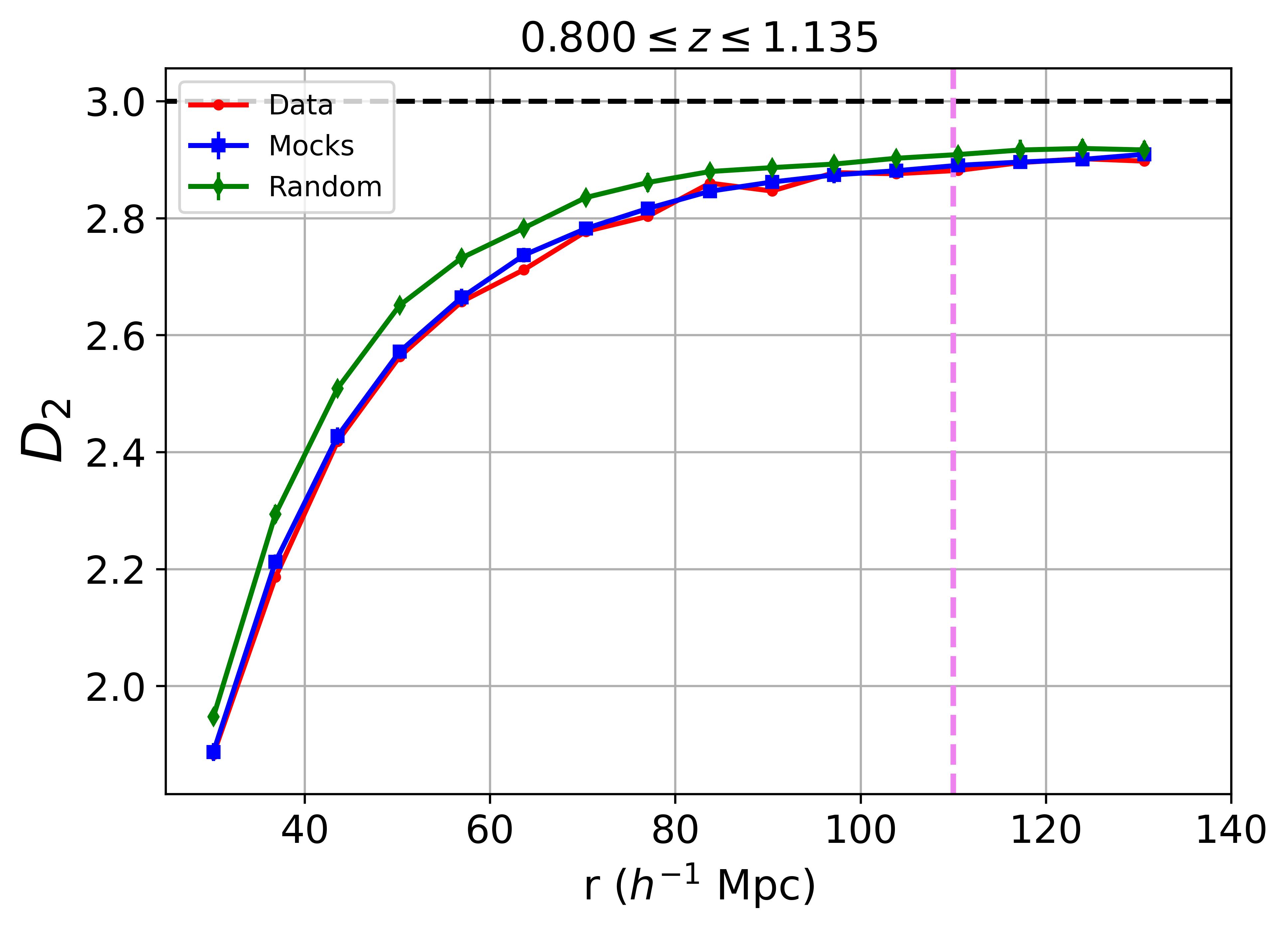}
	\includegraphics[width=0.89\columnwidth]{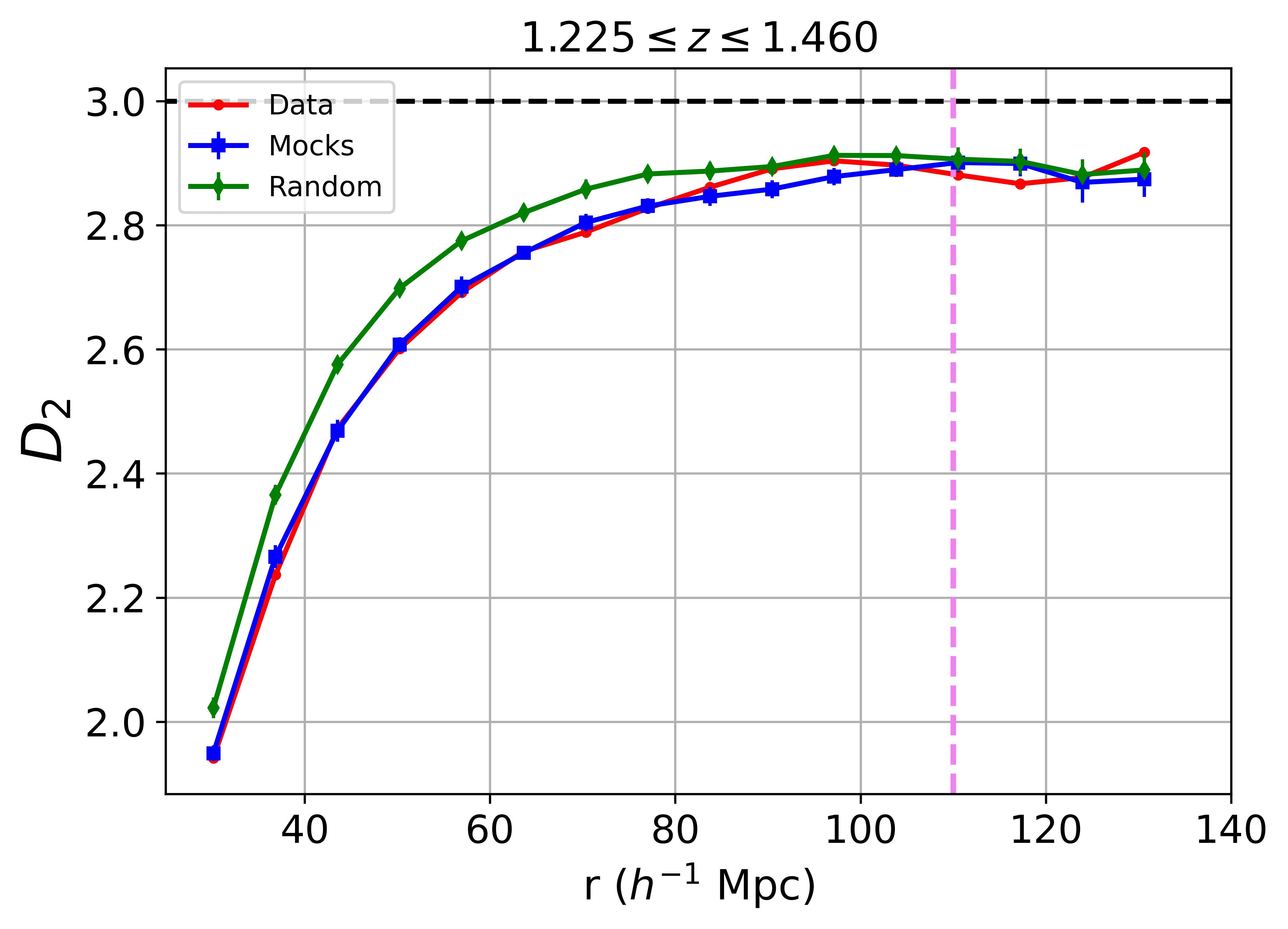}
	\includegraphics[width=0.89\columnwidth]{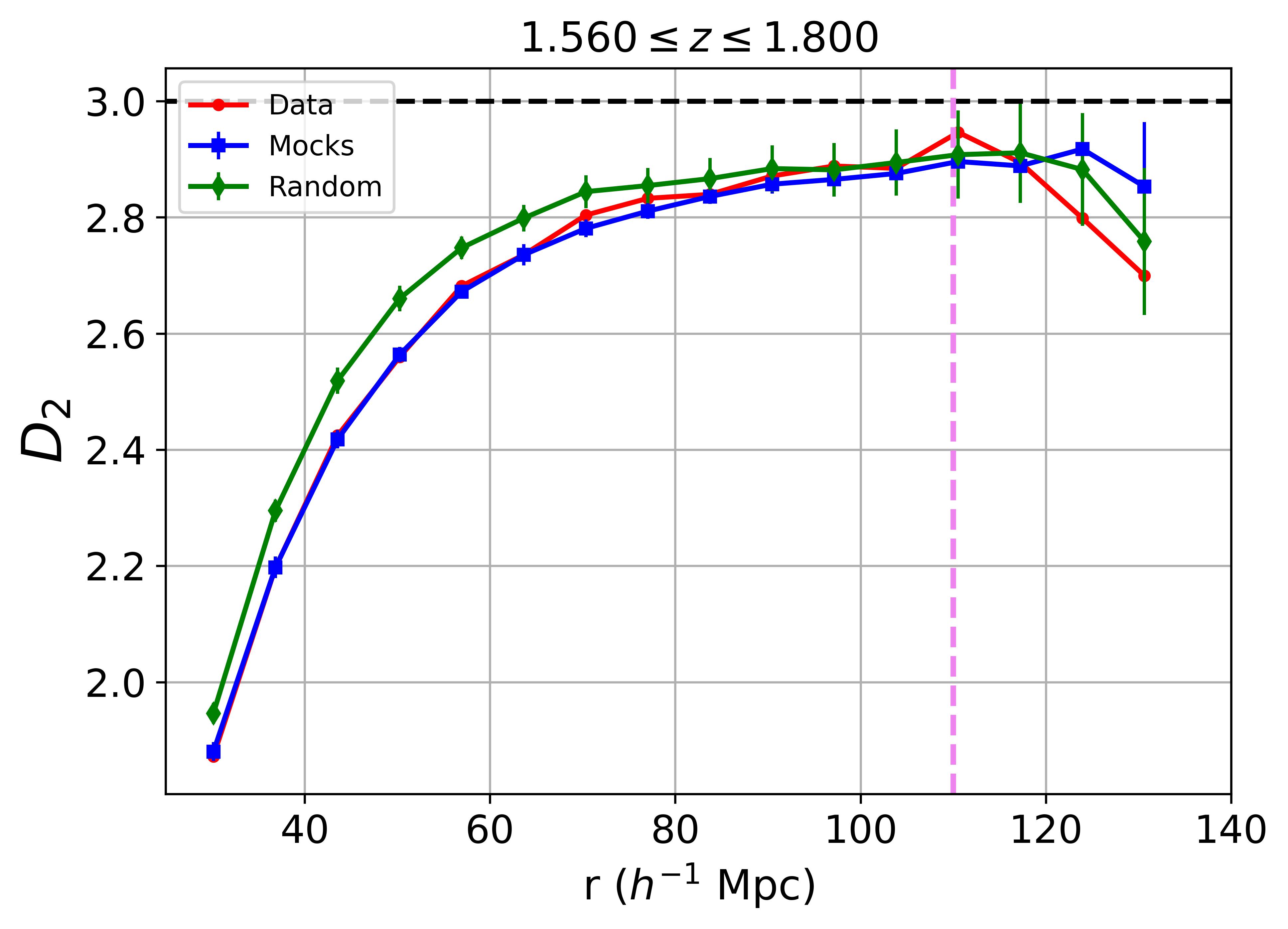}
	\includegraphics[width=0.89\columnwidth]{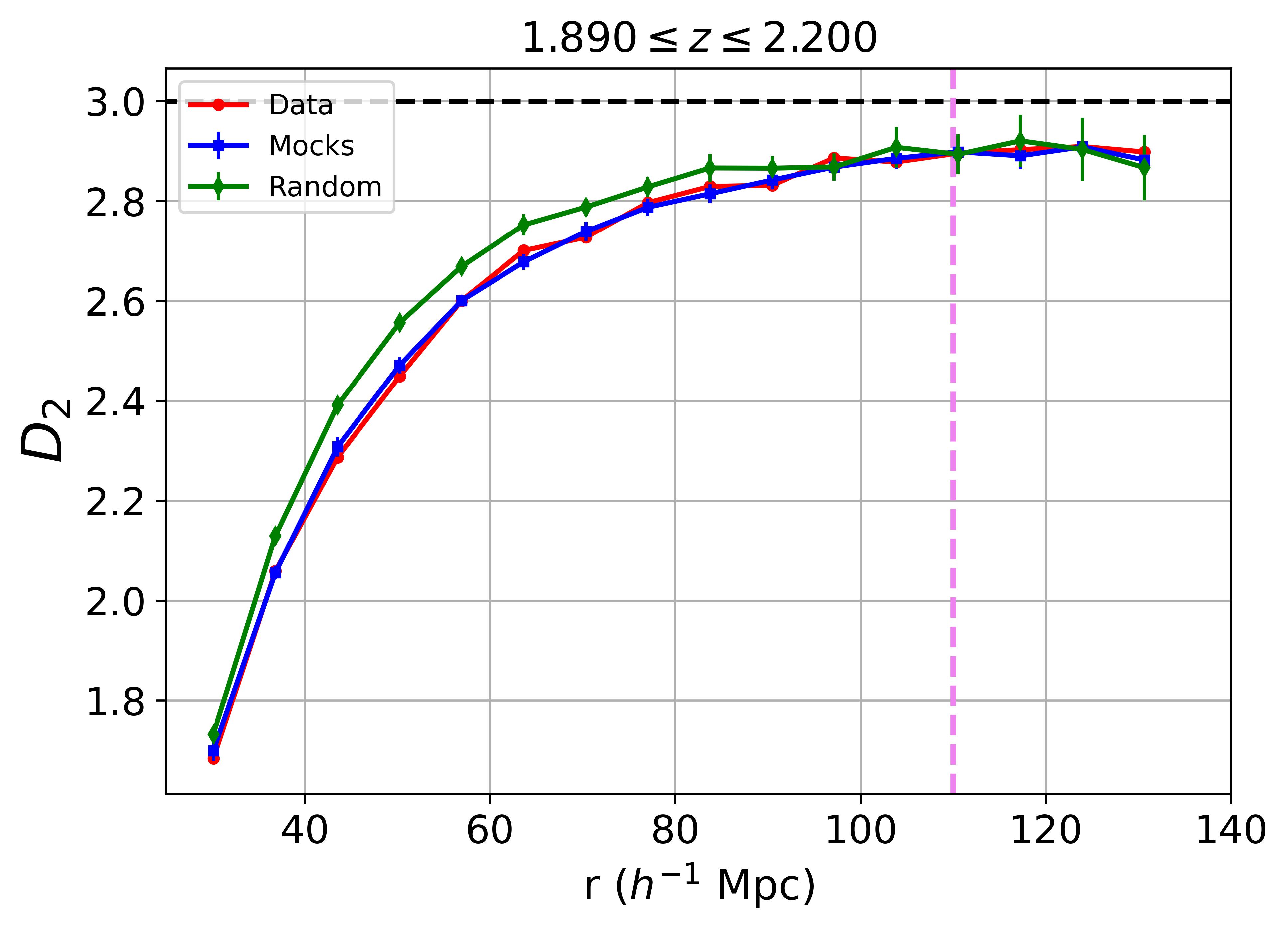}
    \caption{The four panels in this figure shows the Correlation Dimension, $D_q(r)$ for q$=2$  versus distance scale (r) for the observed (red curve) and mock (blue curve) quasar data, and the random (red curve) quasar distribution in the NGC region at each of the four redshift intervals used in our study. The mean and $1\sigma$ error bars for the mock data and the randoms are the mean and sample variance of the 40 realizations of each. The black dashed horizontal line marks $D_2 = 3$. The pink vertical dashed line in each panel at $r_{h} \sim 110$ $h^{-1}$ Mpc represents the scale beyond which each $D_2$ curve saturates.}
    \label{fig:d2_NGC}
\end{figure*}

\begin{figure*}
    \includegraphics[width=0.89\columnwidth]{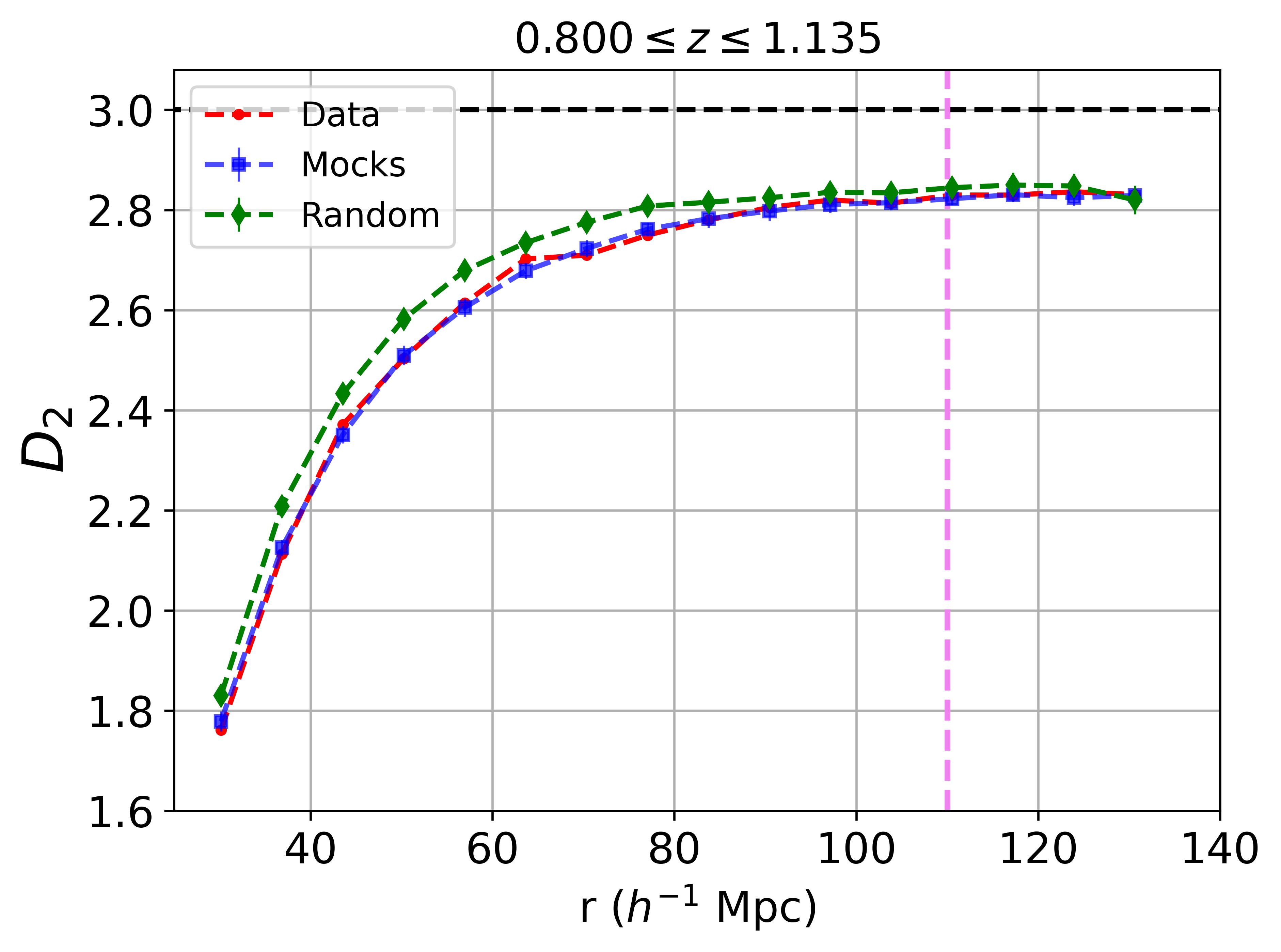}
	\includegraphics[width=0.89\columnwidth]{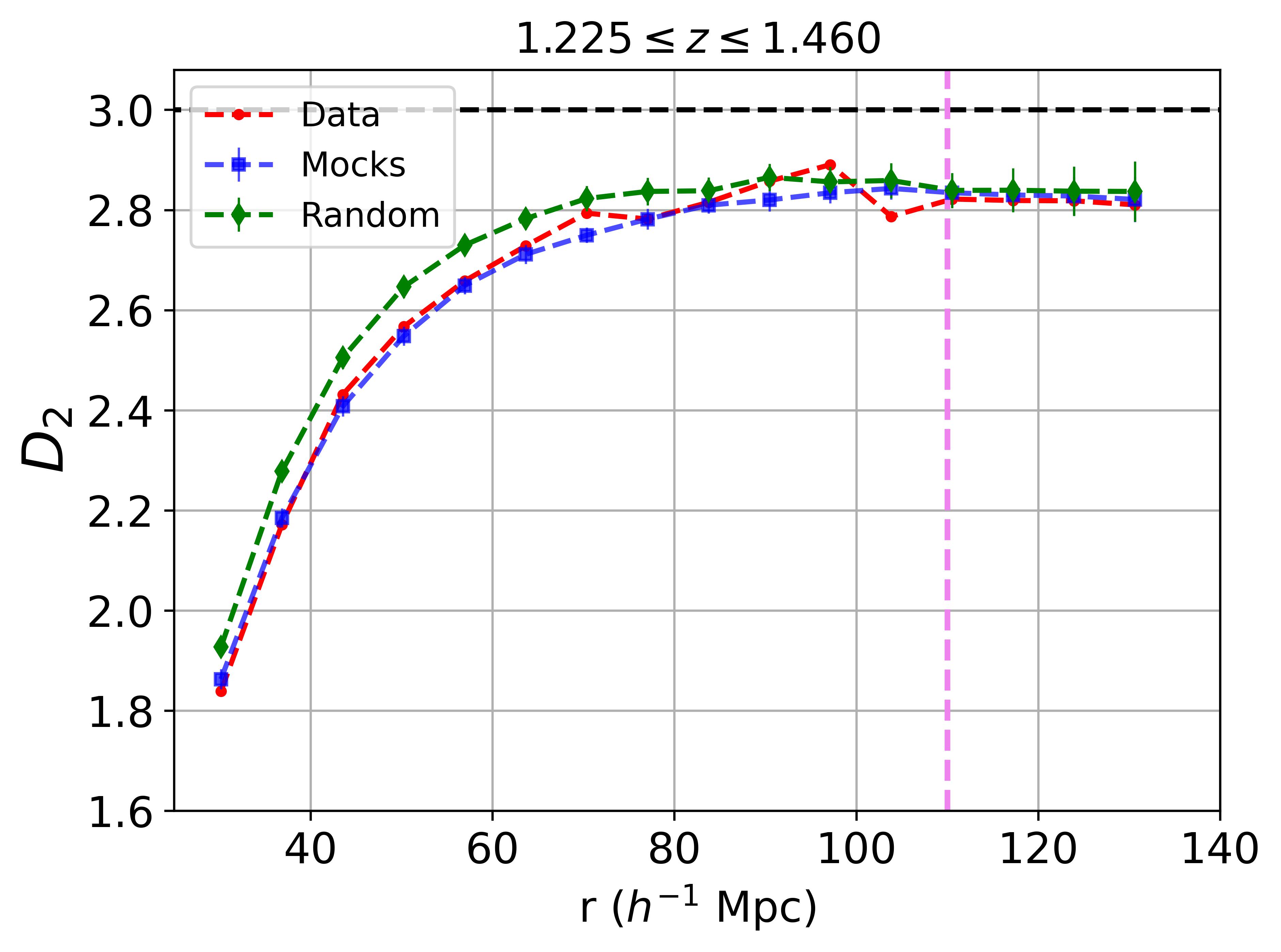}
	\includegraphics[width=0.89\columnwidth]{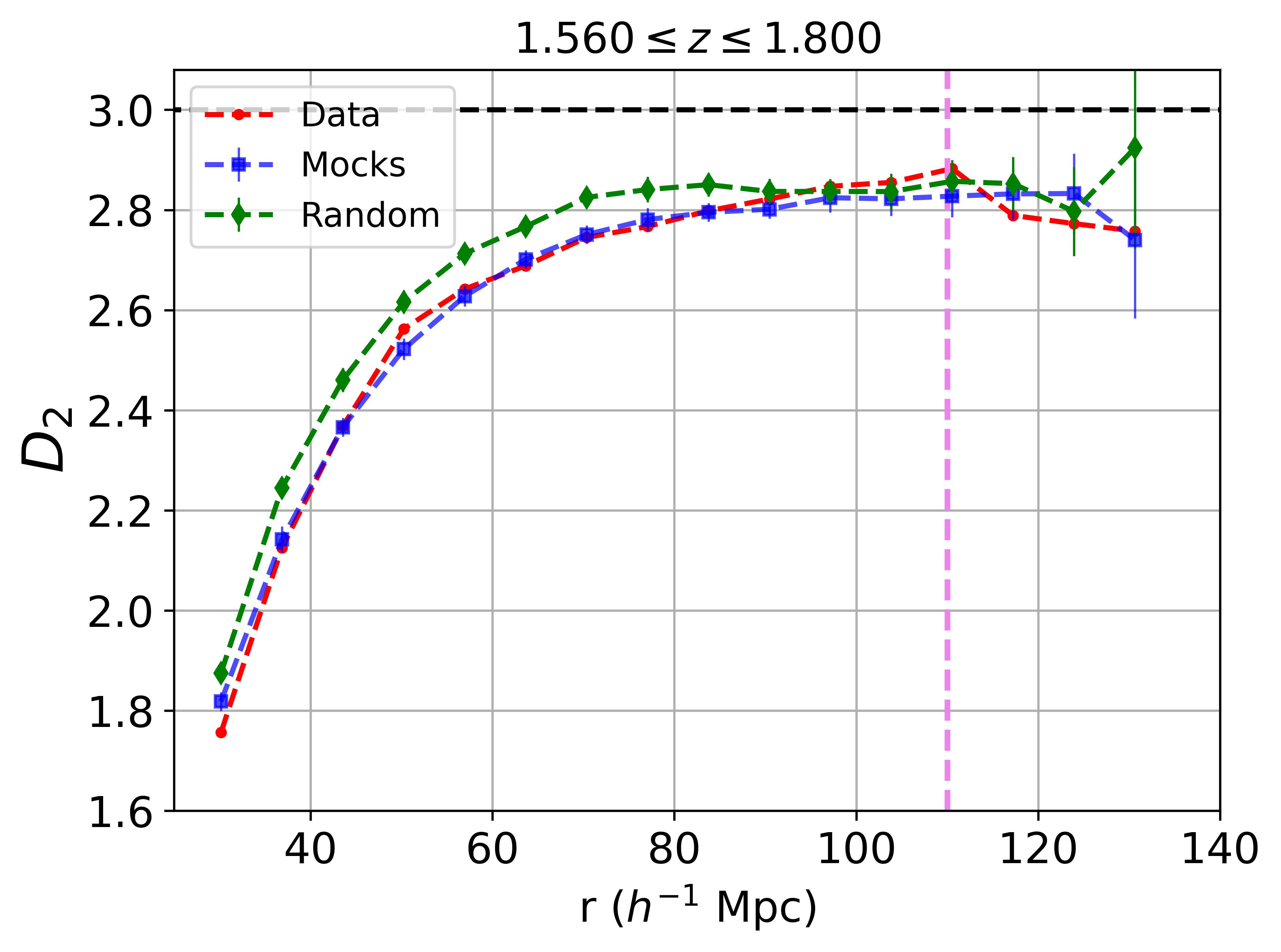}
	\includegraphics[width=0.89\columnwidth]{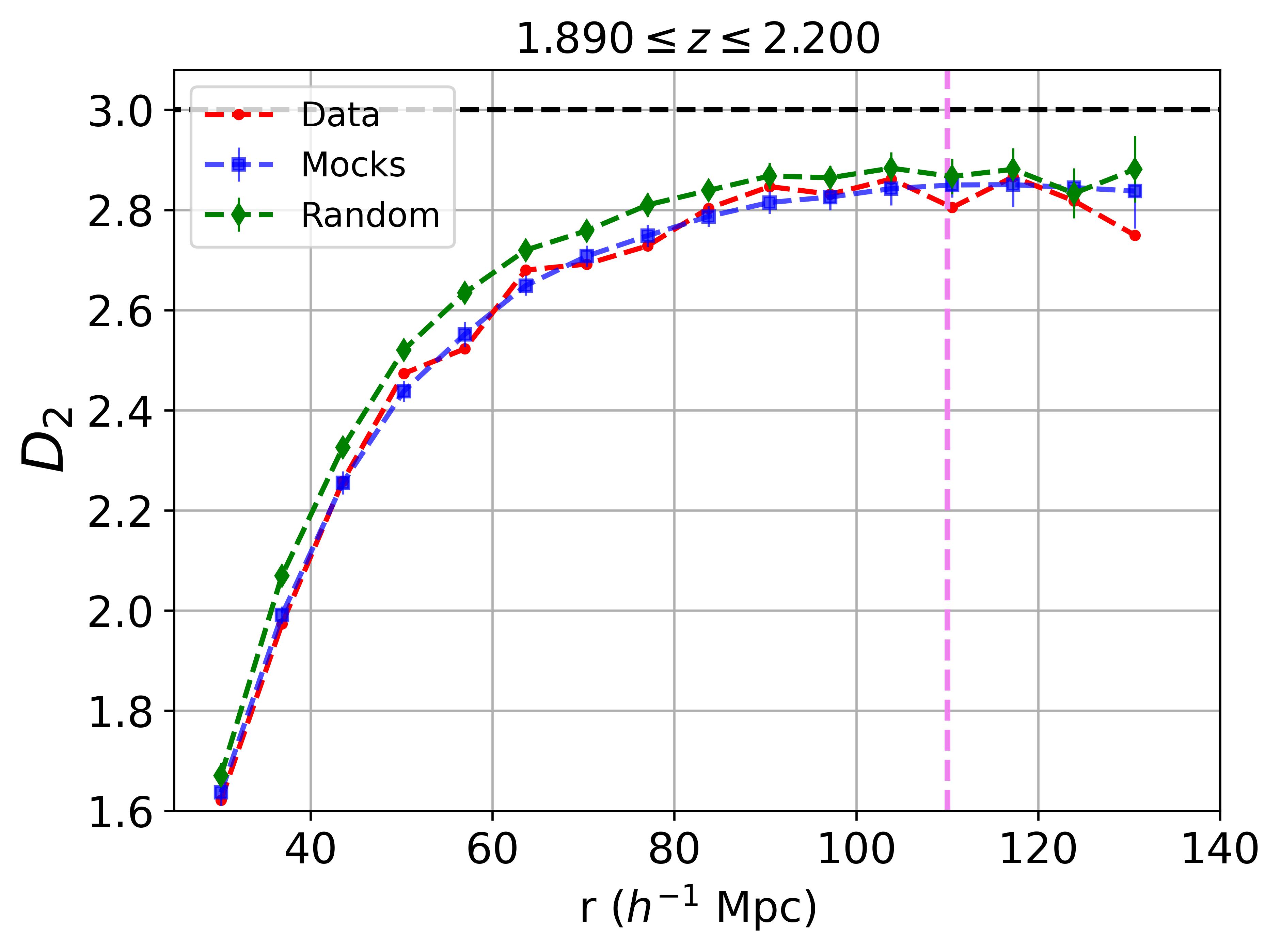}
    \caption{The four panels in this figure shows the Correlation Dimension, $D_q(r)$ for $q=2$  versus distance scale ($r$) for the observed (red curve) and mock (blue curve) quasar data, and the random (red curve) quasar distribution in the SGC region at each of the four redshift intervals used in our study. The mean and $1\sigma$ error bars for the mock data and the randoms are the mean and sample variance of the 40 realizations of each. The black dashed horizontal line marks $D_2 = 3$ for the SGC region. The pink vertical dashed line in each panel at $r_{h} \sim 110$ $h^{-1}$ Mpc represents the transition to the homogeneity scale. }
    \label{fig:d2_SGC}
\end{figure*}

As described above, we have now obtained the correlation dimension, $D_2{(r)}$ defined in Equation~\ref{eqn:Dimension}. 
 The curves in Figure~\ref{fig:d2_NGC} represent the variation of $D_2$ with $r$, obtained for NGC quasar data (red curve), their corresponding mock realizations (blue curve) and the random distribution (green curve), with each panel corresponding to each of the four redshift bins considered in our analysis.
 The mean and the error bars for the mock and random curves are the average and $1\sigma$ sample variance over $40$ realizations. Analogous $D_2$ curves (dash lines) for the quasar distribution in the SGC region are presented in Figure~\ref{fig:d2_SGC}. We find $D_2$ values of the observed quasar data and the simulated mock data to be consistent (within $1\sigma$) with the random distribution beyond the comoving length scale of $r\sim80$ $h^{-1}$ Mpc. However, we notice $D_2$ curves for data and mock deviating from that of the random distribution below $r<80$ $h^{-1}$ Mpc. Also, at lower values of $r$, the $D_2$ curve for random distribution appears to be rising faster than the $D_2$ for the observed data and the simulated data, which is as expected. The $D_2$ for all three distributions under investigation seems to be saturated in the range of 2.8 to 2.9 for $r>80$ $h^{-1}$ Mpc, indicating the transition to homogeneity beyond this length scale. The conventional way adopted in literature to estimate the homogeneity scale ($r_h$) is to consider the scale where the data become consistent with the random distribution within $1\sigma$~\citep{Hogg:2004vw,yadav2005,10.1111/j.1365-2966.2010.16612.x} or to fit a polynomial expression to describe the evolution of each $D_2(r)$ and then find the corresponding homogeneity scale ($r_h$) where $D_2 = 2.97$ (i.e. $1\%$ of the ambient dimension, which is equal to 3 for a distribution in 3-dimensions) for each realization~\citep{Scrimgeour:2012wt}. However, these methods are arbitrary and lack a strong physical motivation. Also, it is important to note that our correlation dimension measure does not attain the value of $D_2 = 3$, which is the ambient integer dimension in our case, even at the largest scales of our analysis. This aligns with earlier studies  ~\citep{2008MNRAS.390..829B,10.1111/j.1365-2966.2010.16612.x}, where it has been demonstrated that in all practical cases of interest, the fractal dimension differs from the ambient integer dimension of the space. These deviations in the fractal dimension are mainly attributed to weak clustering present in the galaxy/quasar distribution, along with a smaller contribution arising due to the finite number of galaxies/quasars present in the distribution. $D_2(r)$ curves for the observed quasar data, mocks, and random distribution in the SGC region as represented in Figure~\ref{fig:d2_SGC} exhibits similar behavior. We do not notice any specific trend with respect to redshift in our correlation dimension analysis. To glean higher-order clustering information from both overdense (positive q moments) and underdense (negative q moments) regions of matter density as traced by the quasar distribution, we extend the correlation dimension to the spectrum of generalized dimension, $D_q(r)$ in the next section.

\begin{figure*}
	\includegraphics[width=\columnwidth]{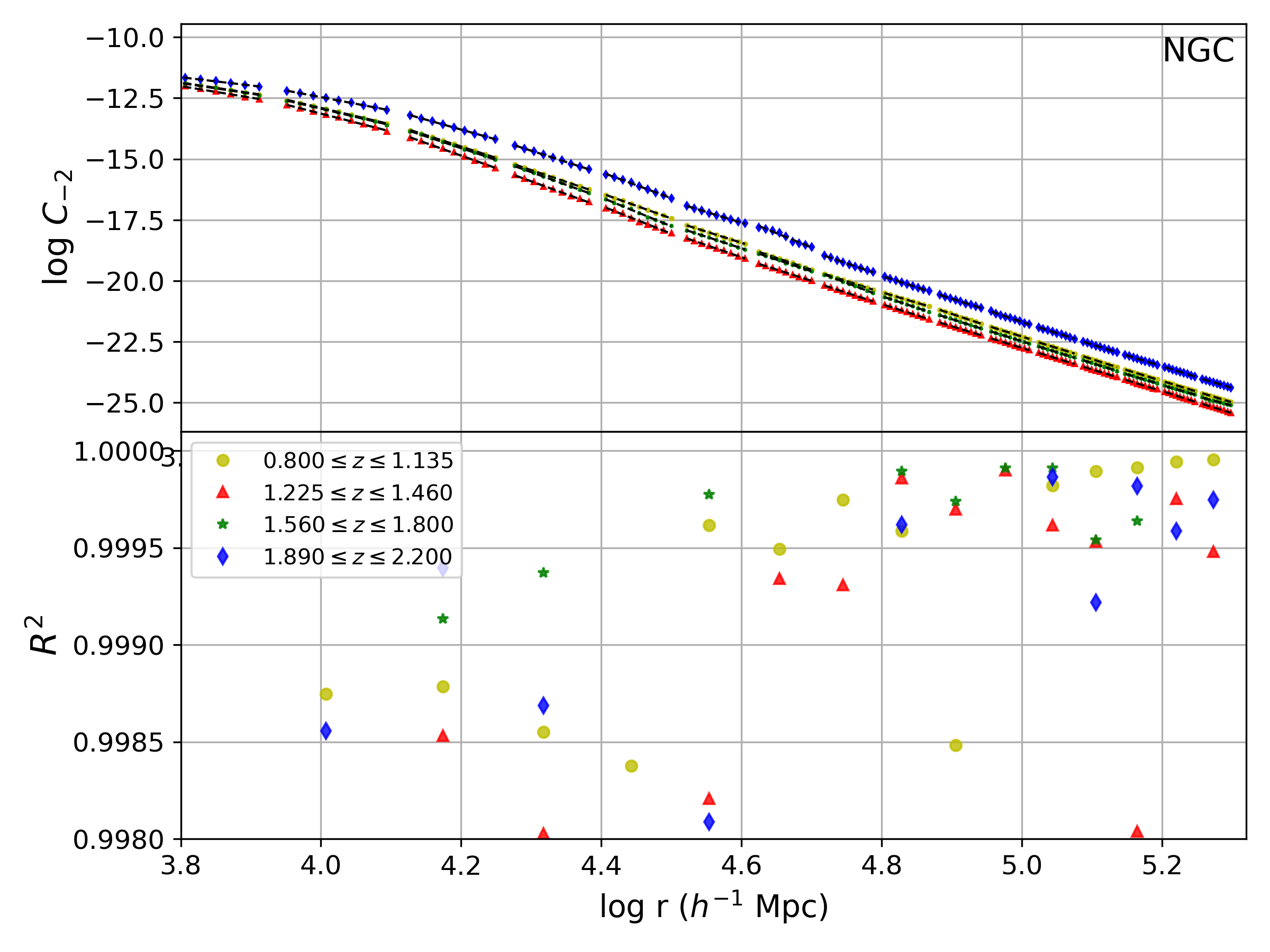}
    \includegraphics[width=\columnwidth]{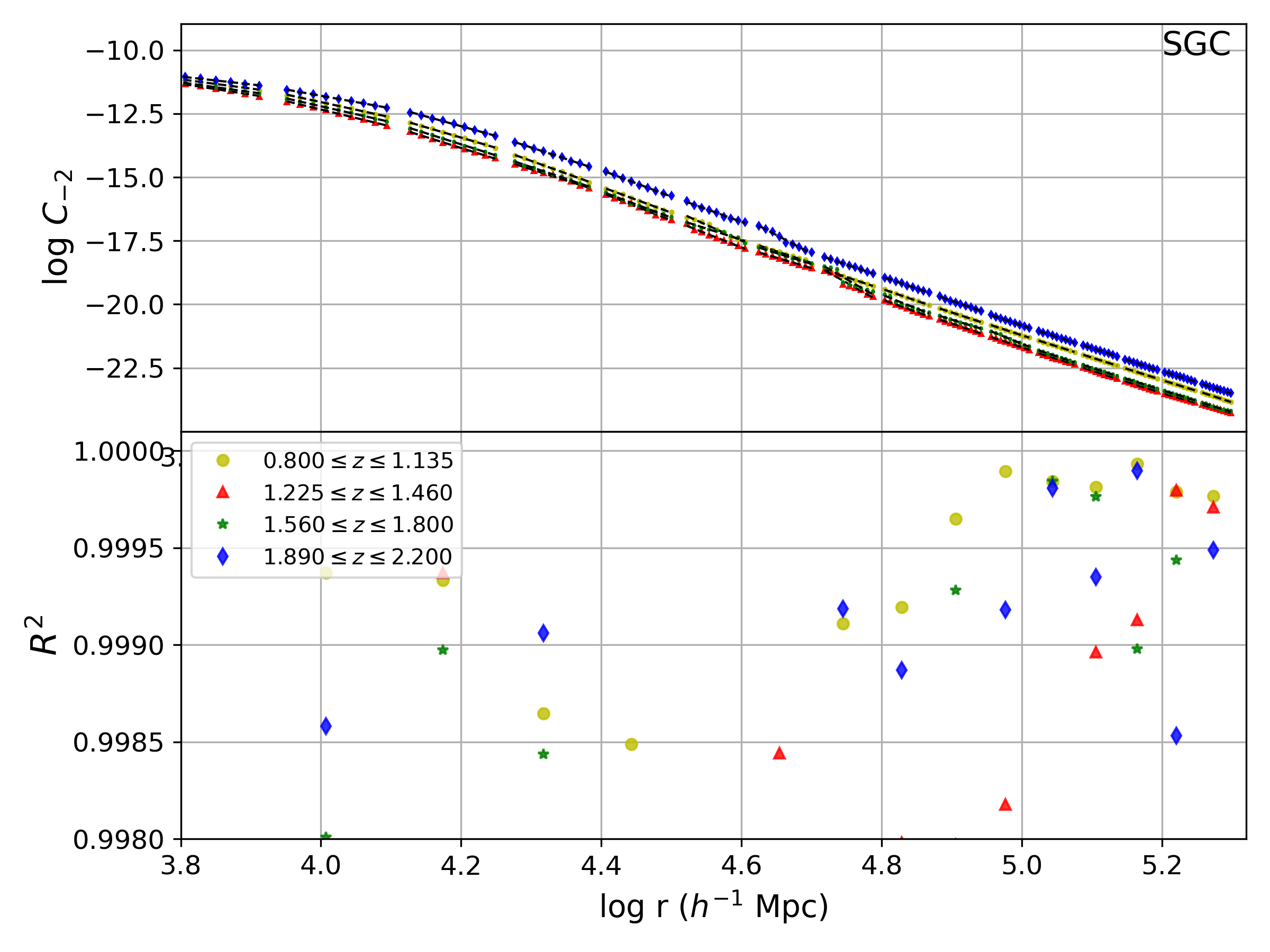}
\caption{This plot depicts log $C_q (r)$  versus log r for q$= -2$, for the observed quasar data in both the NGC (upper left panel) and SGC region (upper right panel). Each curve corresponds to one of the four redshift bins with mean redshifts of $\bar{z}=0.967$ (yellow circles), $\bar{z}=1.342$ (red triangles), $\bar{z}=1.680$ (green stars), and $\bar{z}=2.045$ (blue diamonds). Below the data points we also plot the best-fit line shown in black across different length scales.  This clearly reveals the power law scaling behaviour of $C_{-2}(r)$ with a negative exponent across different length scales. Furthermore, in the lower panels of each plot, we illustrate the goodness-of-fit parameter ($R^2$) as a function of scale. This metric quantifies the robustness of our linear fitting. Remarkably, its value exceeds 0.998 across all length scales, indicating the validity of the assumption regarding the power law behavior of $C_{-2}$.}
    \label{fig:C2_SGC}
\end{figure*}
\begin{figure*}
    \includegraphics[width=0.89\columnwidth]{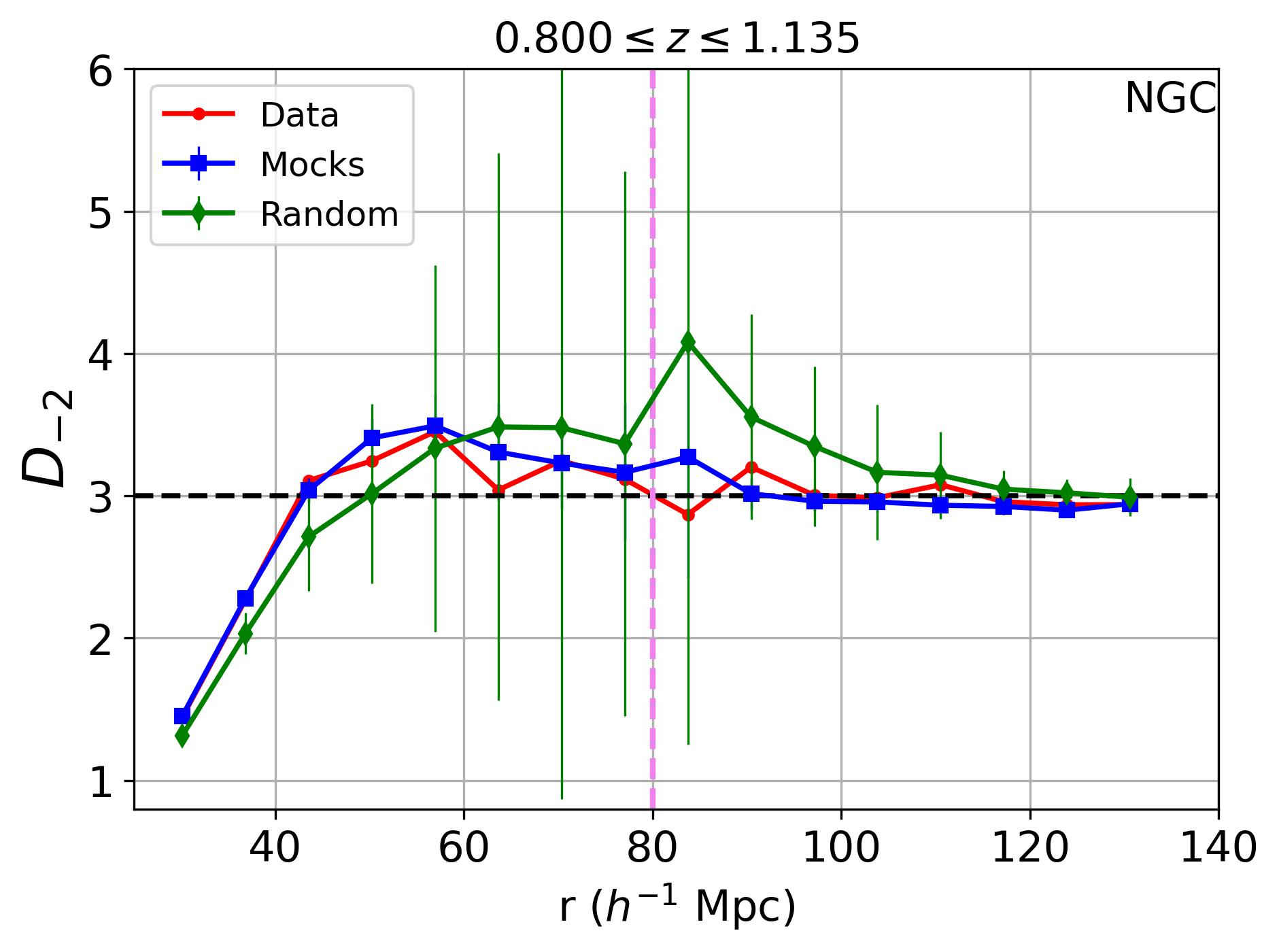}
	\includegraphics[width=0.89\columnwidth]{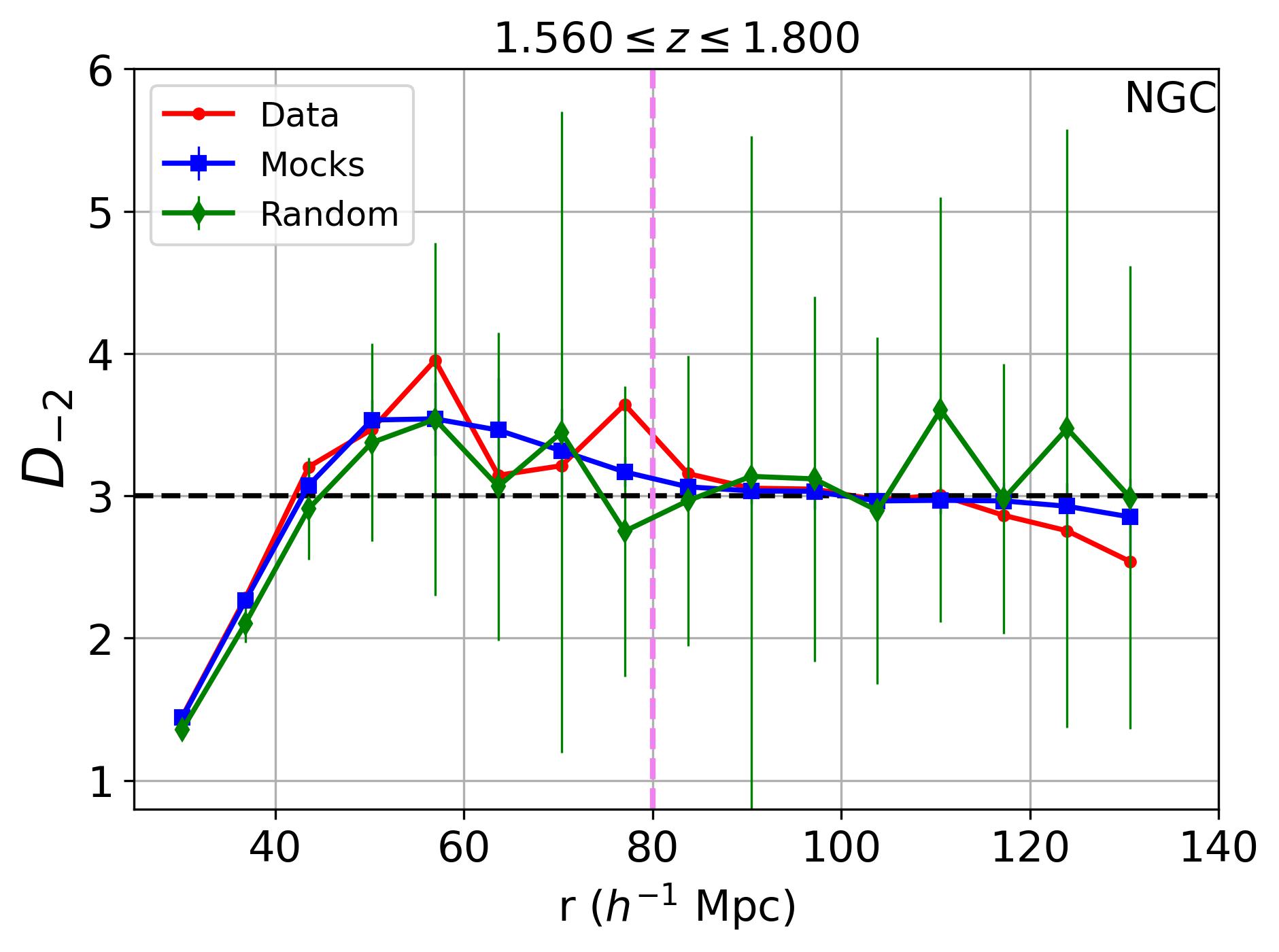}
	\includegraphics[width=0.89\columnwidth]{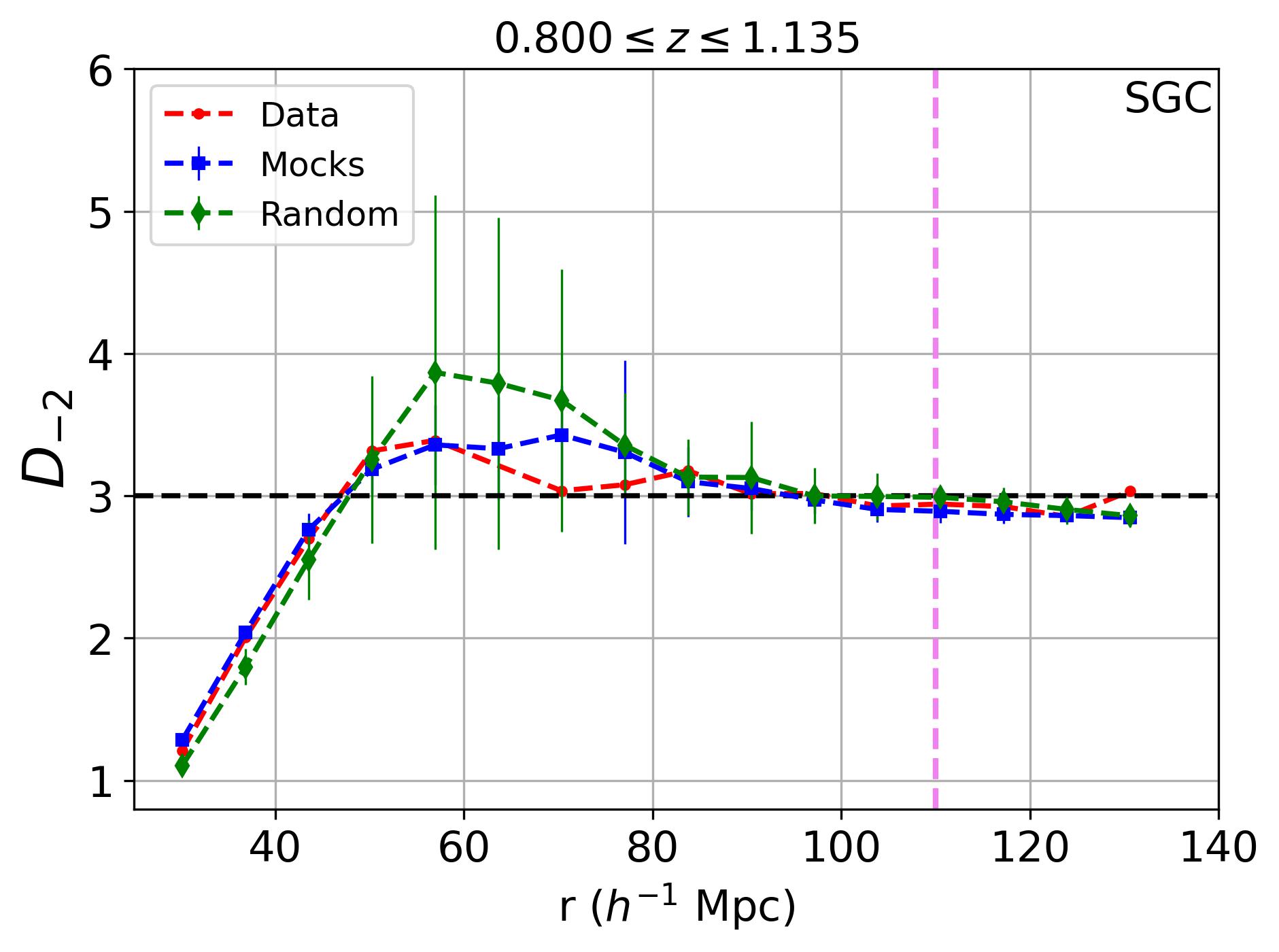}
	\includegraphics[width=0.89\columnwidth]{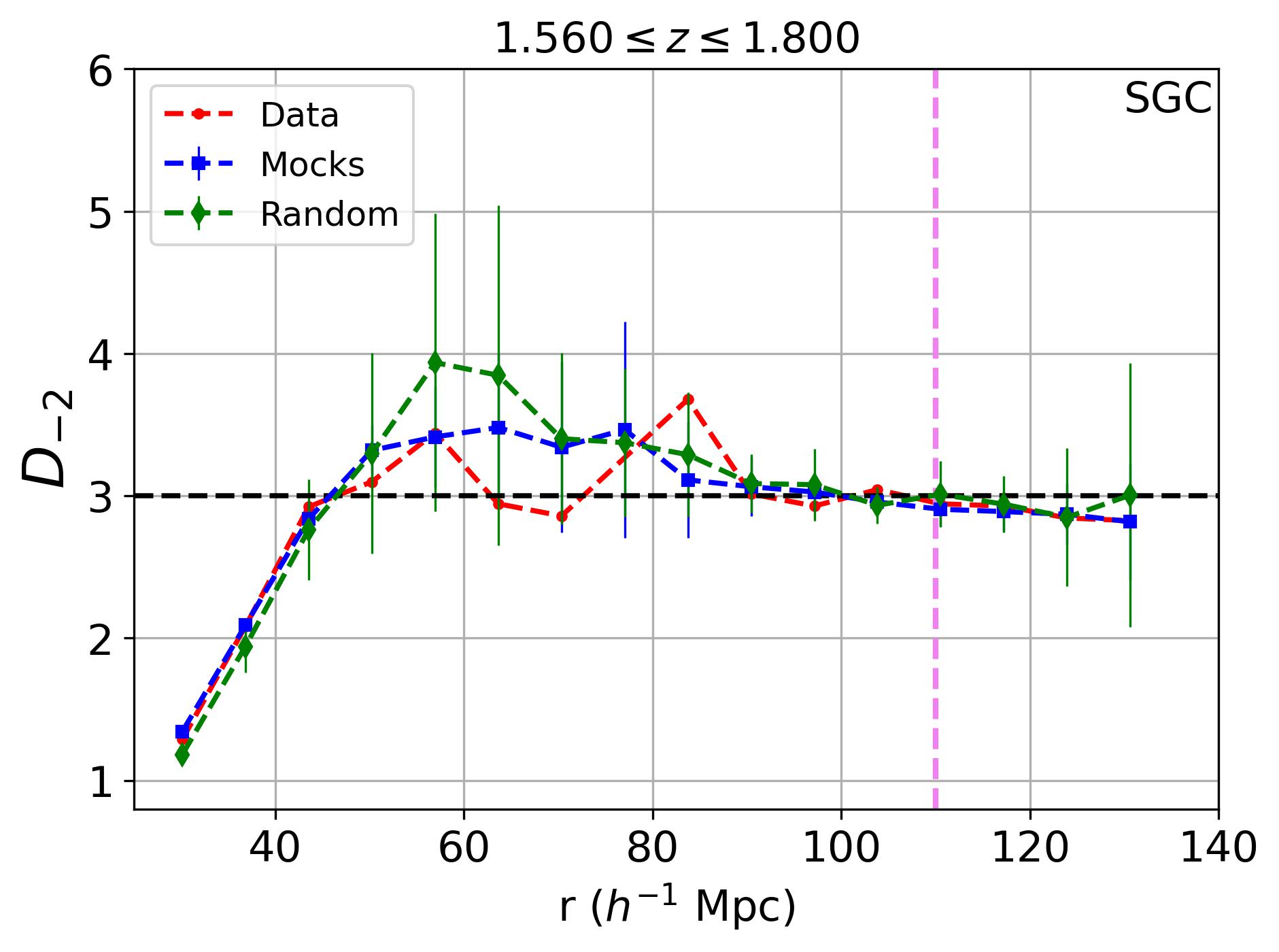}
    \caption{The four panels in this figure shows $D_q(r)$ for $q= -2$ versus comoving radius ($r$) (of sphere) for the observed (red curve) and mock (blue curve) quasar data, and the random (red curve) quasar distribution in the NGC region (top two panels) and SGC region (bottom two panels) in two redshift intervals with mean $z=0.967$ and $1.680$. The mean and $1\sigma$ error bars for mock data and the randoms are the mean and sample variance over the 40 realizations of each. The black dashed horizontal line marks $D_2 = 3$. The pink vertical dashed line in each panel at $r_{h} \sim 110$ $h^{-1}$ Mpc represents the transition to homogeneity scale.}
    \label{fig:dm2_NGC}
\end{figure*}
\begin{figure*}
	\includegraphics[width=\columnwidth]{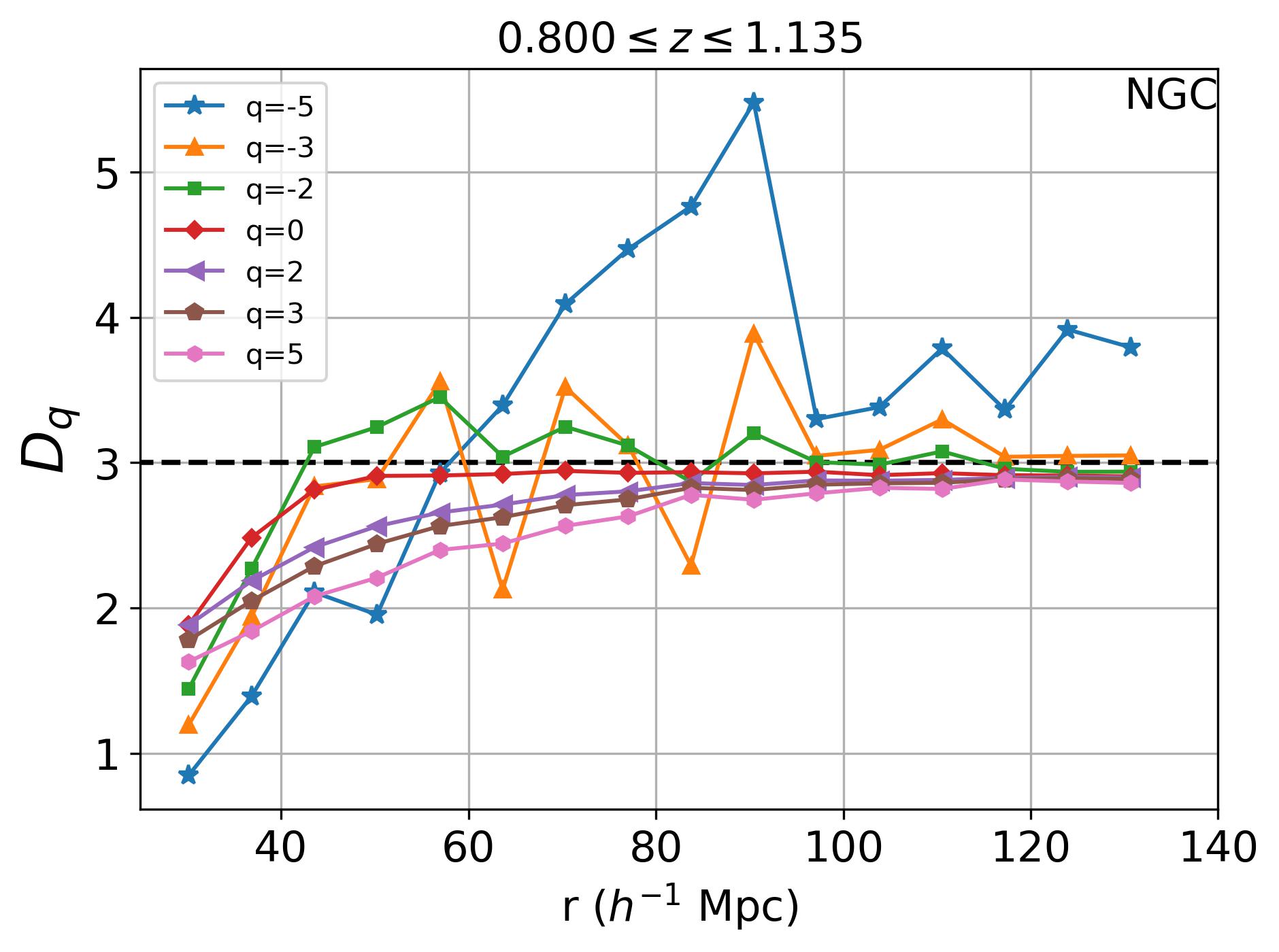}
    \includegraphics[width=\columnwidth]{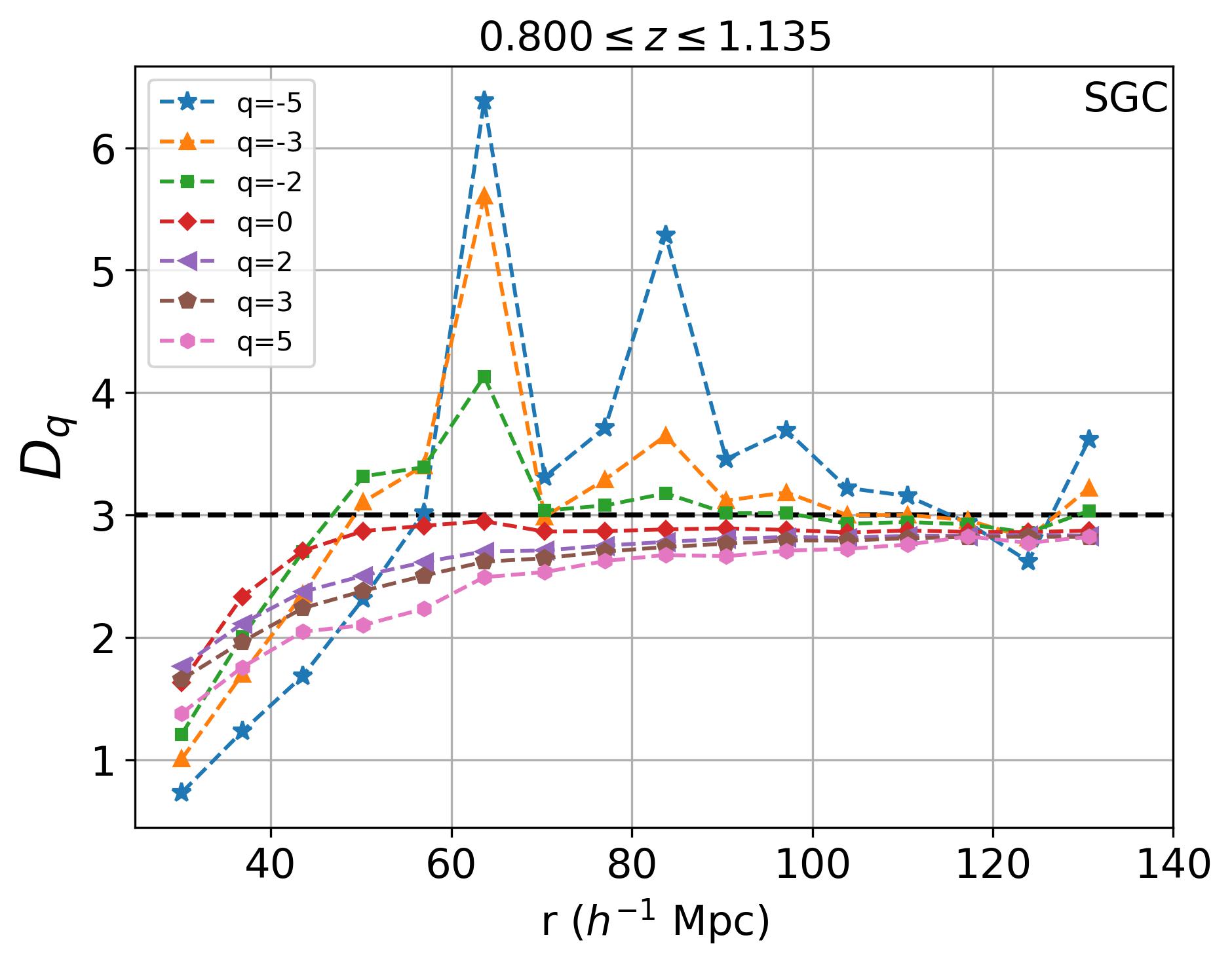}
\caption{The figure shows the variation of  $D_q(r)$ with r, for the observed quasar data in the first redshift interval with mean $\bar{z}=0.967$ in NGC ( {\em left panel} ) \& SGC ( {\em right panel} ) region respectively. Each curve represents a different q value explored in our analysis.}
    \label{fig:3d}
\end{figure*}
\subsection{Generalized Dimension} 
To obtain more comprehensible and complete statistical information on the clustering properties of quasar distribution,  it is essential to study the generalized spectrum of the Minkowski Bouligand dimension, i.e., $D_q$. The $q^{th}$ order fractal dimension will contain information up to $(q-1)^{th}$ order correlation function, which is certainly a better estimator of clustering than simply the two-point correlation function \citep{1995PhR...251....1B}. 
We therefore compute the generalised integral $C_q (r)$  as defined in Equation~\ref{eqn:multi} for various positive and negative values of $q$ $\in$ \{-5,-3, -2, 0, 2, 3, 5\}. 
The Figure~\ref{fig:C2_SGC} illustrates the variation of log $C_q$ for $q=-2$ with log $r$ for each of the four redshift bins defined for quasar distribution in both NGC (left plot) and SGC (right plot) region. From these plots, we discern the linear relationship between log $C_{-2}$ and log $r$ with a negative slope.  
We analyse the scaling behaviour of $C_q(r)$ for other positive and negative q values. We find that $C_q(r)$ increases (decreases) monotonically as a power law in $r$ for positive (negative) values of q. However, the exponent of the power law in $r$ varies across the range of $r$ investigated in our study. 
We further determine the generalized dimension $D_q(r)$ as defined in Equation~\ref{eqn:muti_dimension} from log $C_q(r)$ vs log $r$ by performing a liner fitting in the identified power-law regions. We plot this best-fit linear model (as a black line) along with the data points in the top panels of Figure~\ref{fig:C2_SGC}. Additionally, we also plot the $R^2$ coefficient in the lower panels of Figure~\ref{fig:C2_SGC} which gives a measure of how well the linear model fits the log $C_{-2}$ versus log $r$. The slope of this best fit model gives the value $D_{-2}$ or $D_q$ for that interval in the r range. We expect $D_q(r)$ to match with the ambient dimension (i.e. 3) for a homogeneous distribution. The random samples are homogeneous by construction and we expect them to have $D_q(r) = 3$ at all r. We show $D_q(r)$ as a function of $r$ for $q= -2$ in Figure~\ref{fig:dm2_NGC} for the observed quasar data (red curves), mock data (blue curves) and the random distribution (green curves) in the 1st and 3rd redshift bin (defined in Table~\ref{tab:list}) for the NGC (top two panels) and SGC (bottom two panels) region. The $1\sigma$ error bars for the mock data and random sample are obtained by analysing the 40 realizations of each distribution. The $D_{-2}$ curves in Figure~\ref{fig:dm2_NGC} rise at small $r$, with its value reaching higher than the value of ambient dimension and then starts to fall and saturates at a value in the range 2.8 to 2.9 at the large scales. These plots demonstrate the consistency (within $1\sigma$) of the observed and mock data with the random distribution. However, $D_{-2}$ values for all three distributions observed data, mocks, and random exhibits significant fluctuations than their corresponding $D_2$ curves. These fluctuations can be attributed to the sparse distribution of quasars in the low-density regions (traced by negative q moments), leading to higher statistical fluctuations for negative $q$. However, $D_{-2}$ value, converging to a constant value in the range 2.8 to 2.9 for $r>80$ $h^{-1}$ Mpc, again indicates a transition to homogeneity at or beyond this length scale. 
 We illustrate in Figure~\ref{fig:3d}
the behaviour of generalized dimension $D_q(r)$ as a function of $r$ for different $q$ values for the observed quasar distribution in NGC (right panel) and SGC (left panel) region at $\bar{z}=0.967$ (i.e. the first redshift bin). The $D_q(r)$ curves for other values of q exhibit similar behaviour, i.e. $D_q(r)$ rising at small length scales and eventually saturating at higher r values typically beyond $r>80$ $h^{-1}$ Mpc.

\begin{figure*}
  \centering
  \includegraphics[scale=0.48]{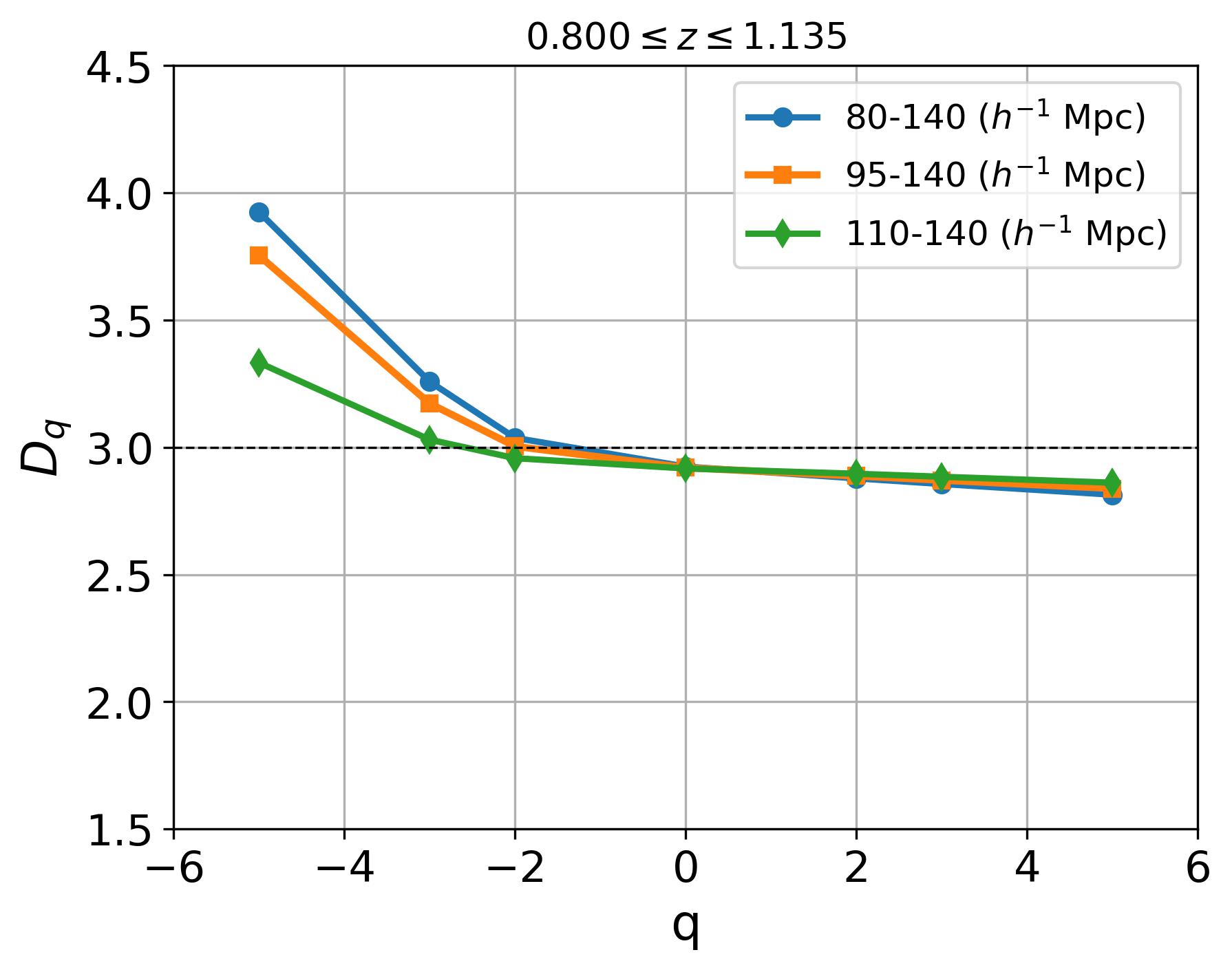}
  \includegraphics[scale=0.48]{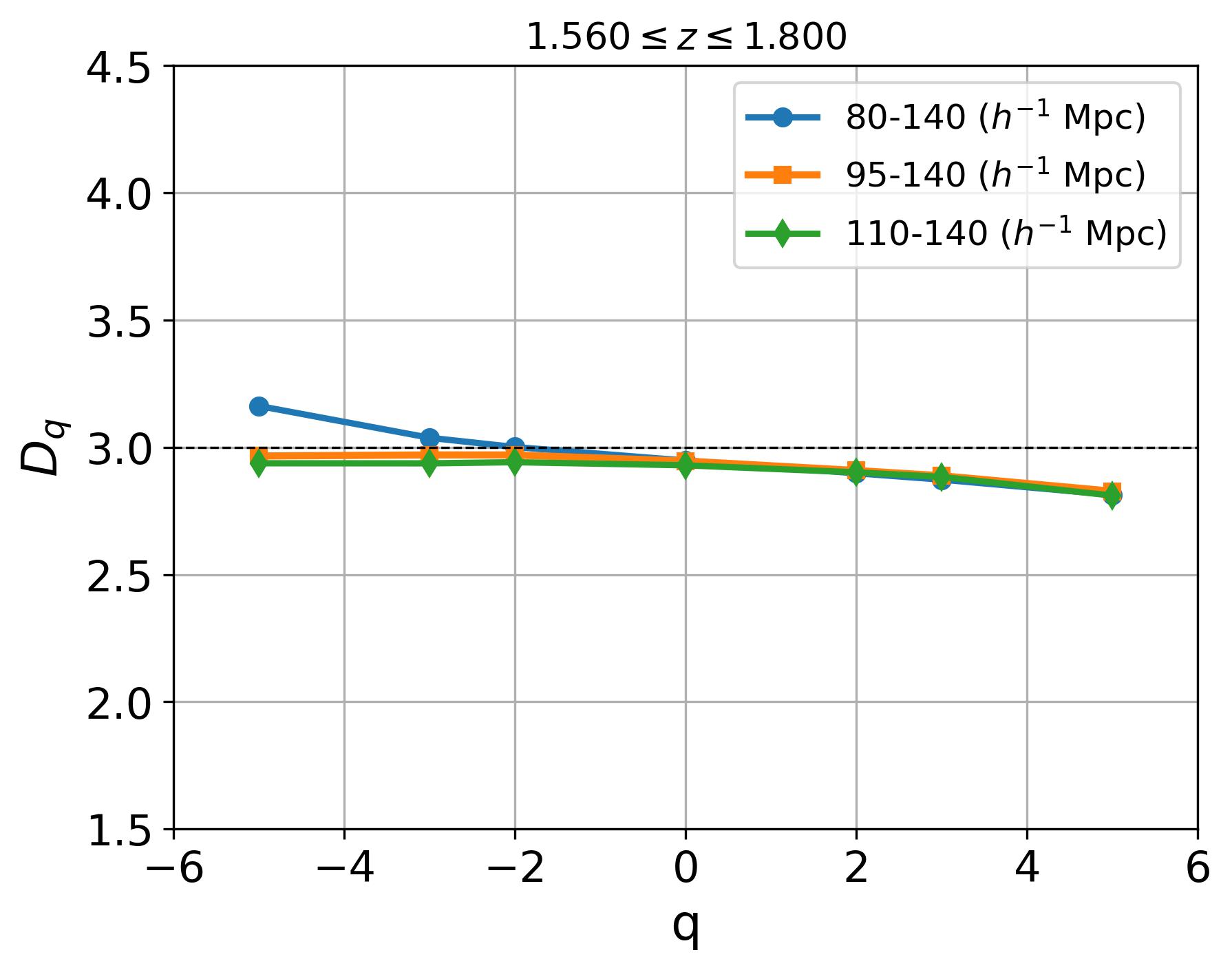}
  \includegraphics[scale=0.48]{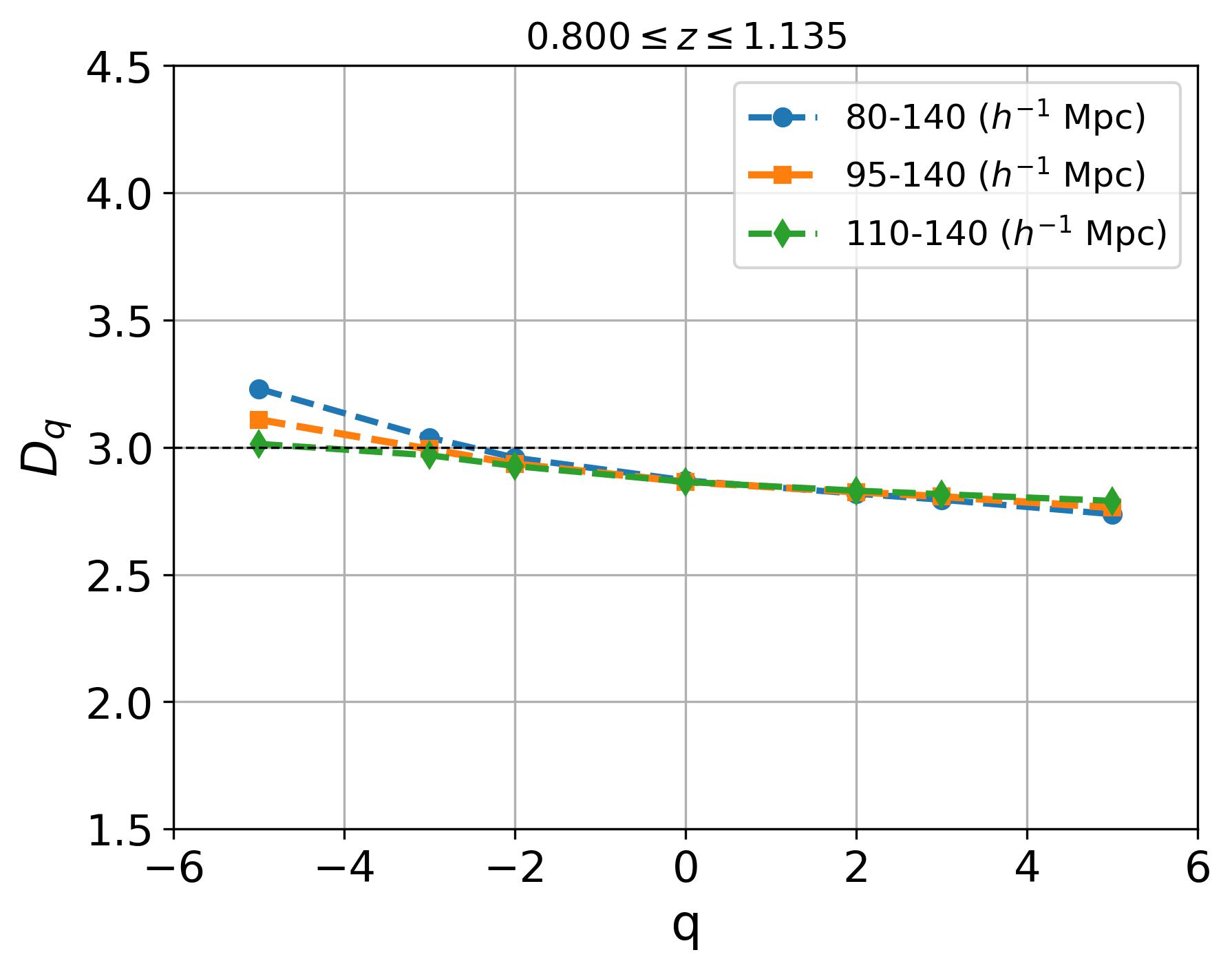}
  \includegraphics[scale=0.48]{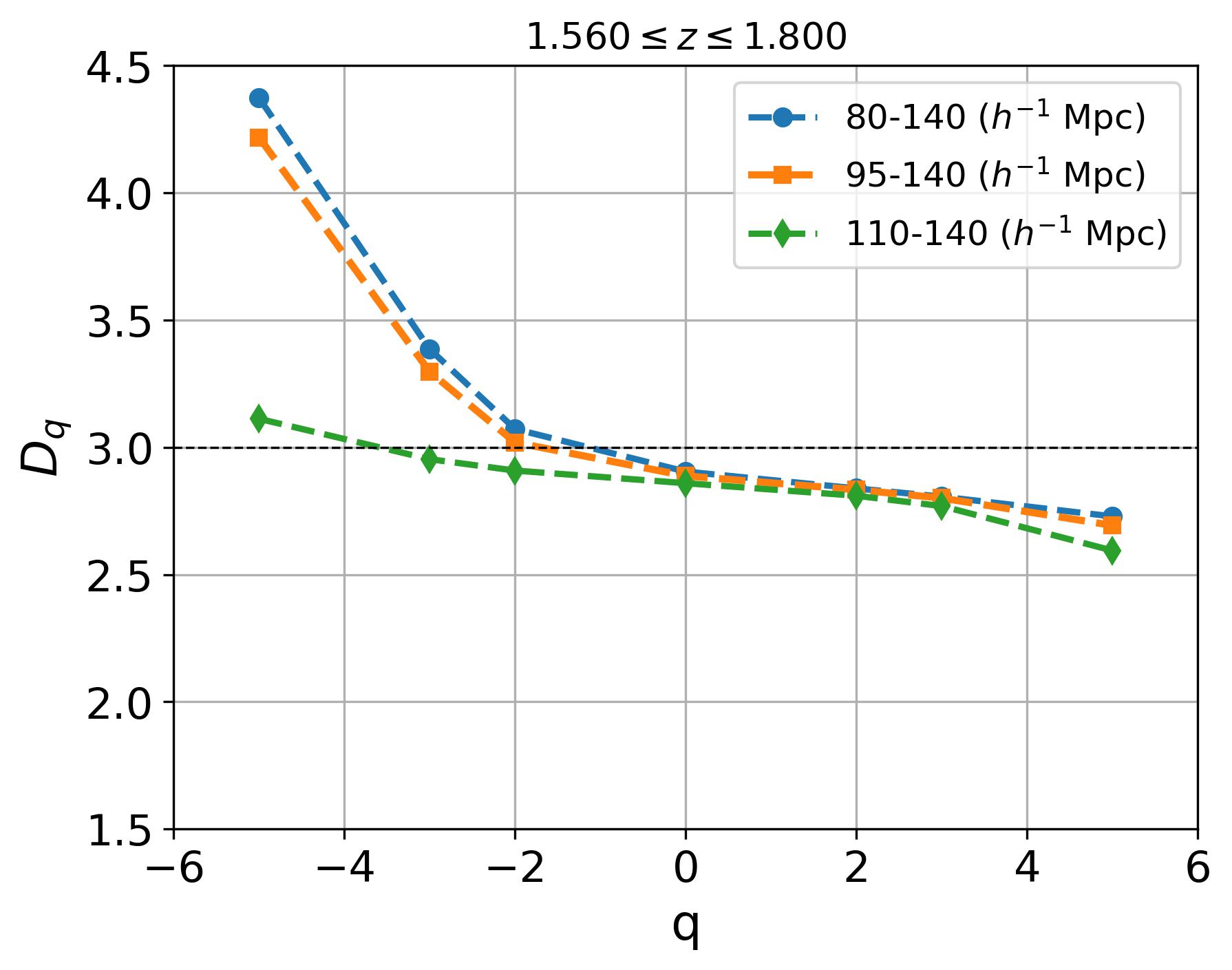}
  
  \caption{{\em Top panels:} These plots show the spectrum of generalized dimensions $D_q$ as a function of $q$ for the observed quasar data in the NGC region investigated across different length scales indicated in the legend of each panel. The two panels correspond to two redshift bins with the mean redshift of $\bar{z}=0.967,  1.680$ respectively. {\em Bottom panels:} These plots also show the variation of $D_q$ with $q$ as in the top panels except for the quasars in the SGC region.} 
  \label{fig:dq_data_NGC}
\end{figure*}

\begin{figure*}
  \centering
  \includegraphics[scale=0.43]{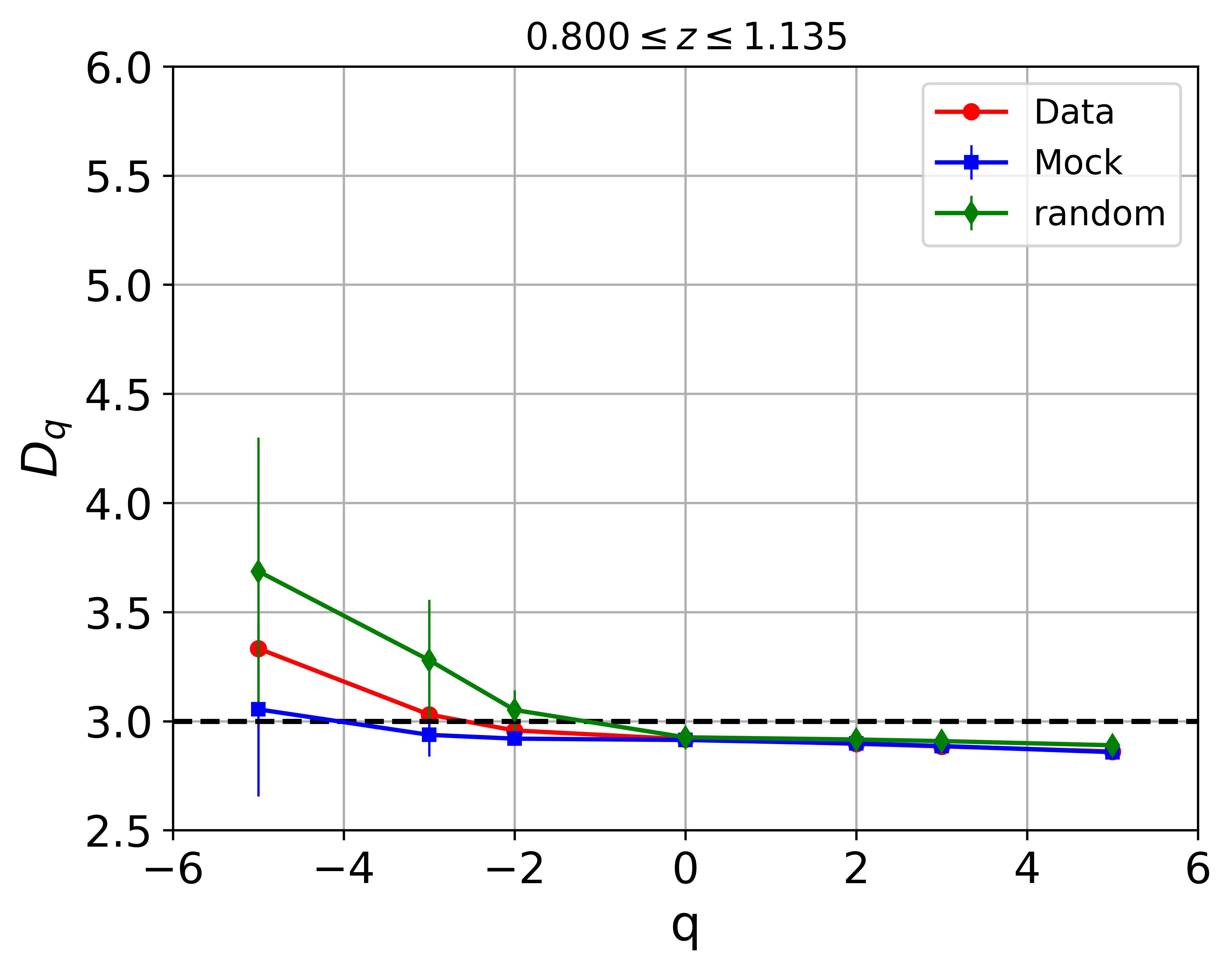}
  \includegraphics[scale=0.43]{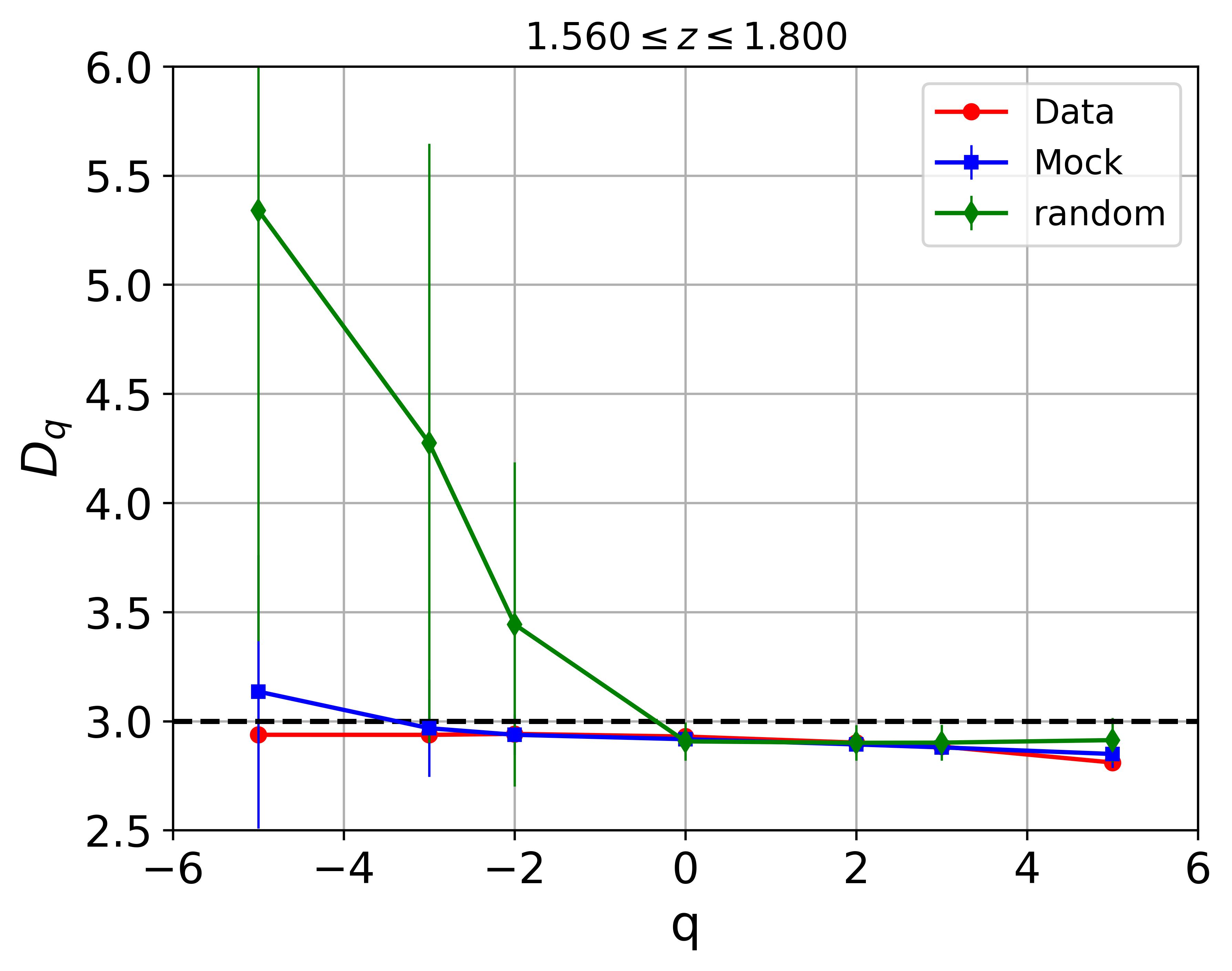}
  \includegraphics[scale=0.43]{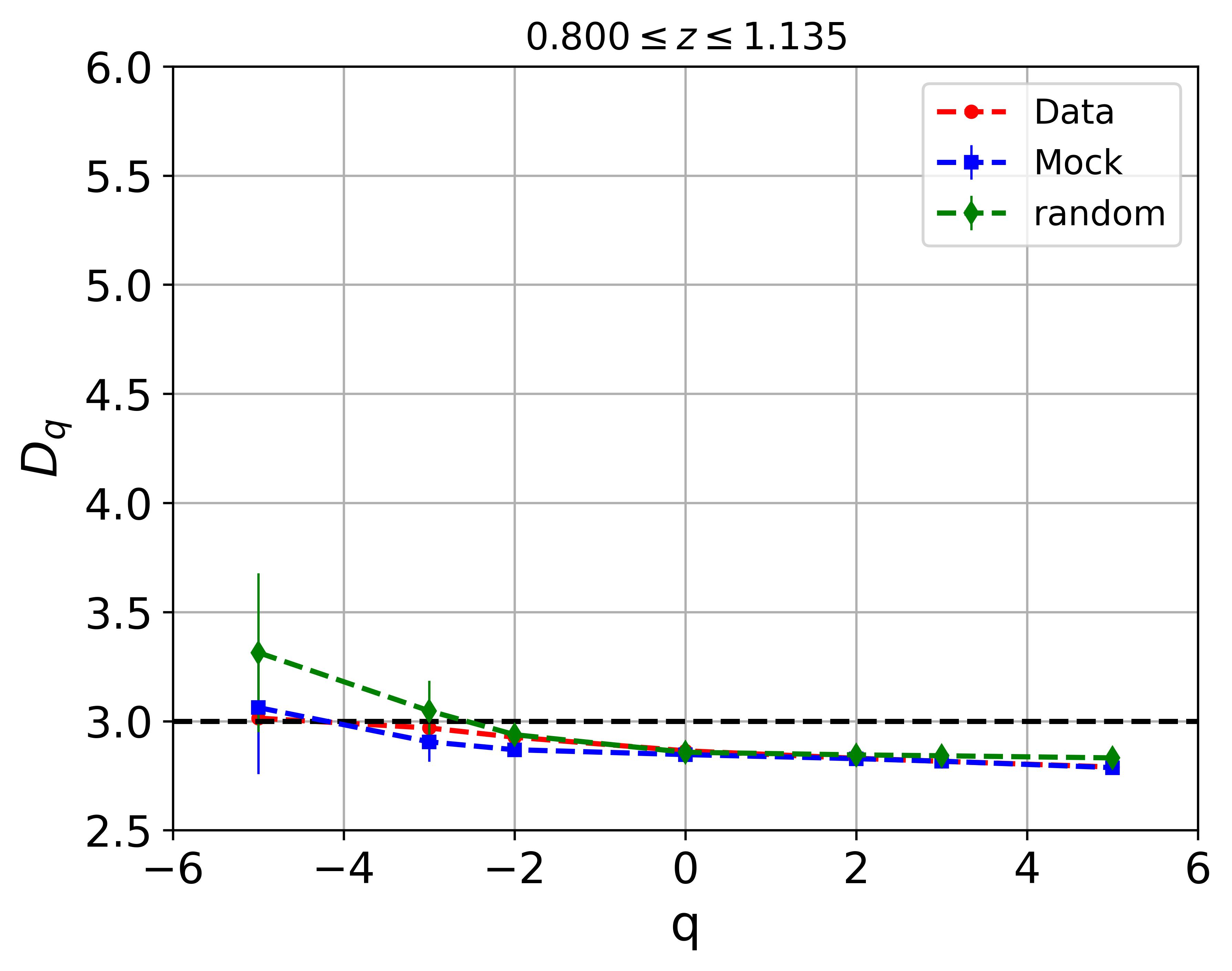}
  \includegraphics[scale=0.43]{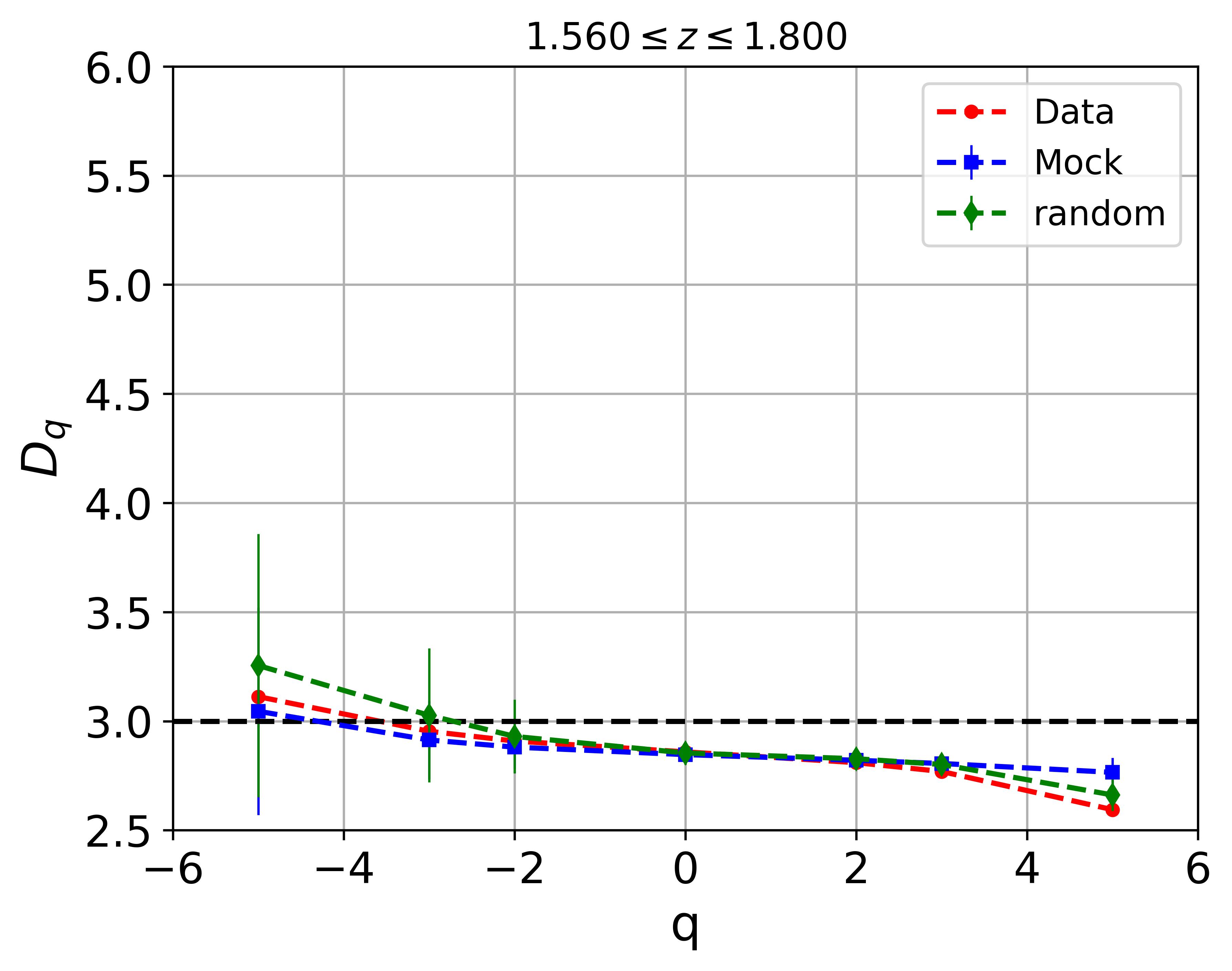}
  \caption{These plots show the spectrum of generalized dimension $D_q$ versus $q$ for the observed data (red curve), the random distribution (green curve) and the mock data (blue curve) between length scales 100 to 140 $h^{-1}$ Mpc. The top two panels correspond to quasar distribution in NGC region for two redshift bins with the mean redshift of $\bar{z}=0.967, 1.680$. The bottom two panels demonstrate the similar variation of $D_q$ versus q for quasars in the SGC region. }
  \label{fig:Dq_NGC}
\end{figure*}

To further extract the transition to homogeneity scale from the Minkowski Bouligand Dimensions analysis, we have computed the $D_q(r)$ values over different ranges of length scales for the observed quasar distribution, the random distribution, and the corresponding mock data, using linear fitting as done before in the case of correlation dimension.
The plots in Figure~\ref{fig:dq_data_NGC} show the $D_q$ versus $q$ curve for the observed quasar data in NGC (top panels) and SGC (bottom panels) regions. The redshift bin is indicated in each panel. The multiple curves in each panel correspond to the different length scales over which scaling behavior is inspected. 
The scaling behavior at small scales ($r<80$ $h^{-1}$ Mpc) differs significantly from that at large scales ($r>80$ $h^{-1}$ Mpc). We analyse and interpret further the large-scale behaviour of the $D_q$. By examining the $D_q$ curves for scales between $80-140$ $h^{-1}$ Mpc, $95-140$ $h^{-1}$ Mpc and $110-140$ $h^{-1}$ Mpc across different redshift bins as displayed in Figure~\ref{fig:dq_data_NGC}, we find $D_q$ curve to be most uniform (and closest to $D_q=3$) across both positive and negative $q$ values in the range $110-140$ $h^{-1}$ Mpc.  We then further checked $D_q$ versus $q$ variation for the corresponding mock data and the random distribution which are shown in Figure~\ref{fig:Dq_NGC} for both NGC (top panels) and SGC (bottom panels) samples. We observe that the $D_q$ curve for the random distribution shows high statistical fluctuations, however, overall we see $1\sigma$ agreement of the observed and mock quasar distributions with the random distribution (homogeneous one). We can thus conclude from our analysis that the Universe exhibits a transition from clustered to being smooth homogeneous beyond the comoving length scale of $r>110$ $h^{-1}$ Mpc. 

\section{Discussion}
\label{discuss}

Our analysis, conducted through multi-fractal examination of the SDSS-IV DR16 eBOSS quasar distribution, corroborates previous findings based on the same dataset ~\citep{Goncalves_2021,Pandey_2021}. Specifically, the measurement of the Baryon Acoustic Oscillation (BAO) feature, which signifies excess clustering amplitude in either the correlation function or the power spectrum, positions it at approximately $100 \ h^{-1}$ Mpc for the SDSS-IV DR16 eBOSS quasar distribution within the redshift range of $0.8 < z < 2.2$~\citep{Hou:2020rse}. Our analysis reveals a transition to homogeneity at a scale surpassing this BAO peak. Therefore, our findings do not contradict the presence of this physical length scale embedded in the quasar distribution.

However, it is noteworthy that several studies~\citep{Hogg:2004vw,yadav2005,Scrimgeour:2012wt,Pandey_2021} have observed that the scale of homogeneity, when deduced from galaxy distribution, tends to be marginally smaller than what our work reports using the quasar distribution. This discrepancy may stem from the potential underestimation of the homogeneity scale in smaller galaxy samples due to the suppression of inhomogeneities resulting from the overlapping of measuring spheres~\citep{Pandey_2021}.

The generalized fractal dimension $D_q(r)$
for the observed distribution of quasars, demonstrate a close agreement (within $1\sigma$) with mock quasar data~\citep{Zhao:2021ahg} generated through N-body simulations based on the Friedmann-Robertson-Walker (FRW) model of the $\Lambda CDM$ cosmology. However, it is crucial to approach these findings with caution when assessing their consistency with the $\Lambda CDM$ model. This caution arises from their inherent dependence on the assumption of the FRW metric, which in turn relies on assumptions of homogeneity and isotropy. To mitigate these potential biases, an alternative approach involves measuring the angular homogeneity scale within the quasar/galaxy distribution. Initially proposed by ~\cite{2014MNRAS.440...10A}, this method has since been applied in various studies~\citep{Goncalves:2017dzs,2022JCAP...10..088A,2022JCAP...04..044C} to test the cosmic homogeneity in a model-independent way. 

The two-point correlation function and consequently, the correlation integral denoted by $C_2(r)$ in this work is expected to be boosted by the “Kaiser factor”. However, this multiplicative amplification factor does not affect our results as the correlation goes to very small values on the length scales of interest in this analysis. Additionally, given that quasars are a highly biased dataset, characterized by a high bias value, that suppresses the amplification factor (Kaiser, 1987).~\citep{1987MNRAS.227....1K}. While a correction based on the quasar bias is expected to offer better insights into the redshift evolution of the universe's homogeneity scale~\citep{Goncalves:2018sxa,Goncalves_2021}, we do not apply bias corrections to $D_q (r)$
calculated in our study. It will be considered in future works. Ultimately, despite quasars being inherently biased tracers of the matter density field, their advantage lies in their ability to sample a vast volume of space. This characteristic makes quasars particularly suitable for testing homogeneity over large scales.

\section{Conclusions}
\label{conclusion}
In our study, we examined cosmic homogeneity on large scales in the distribution of quasars using data from the SDSS-IV DR16 eBOSS survey. This dataset was divided into NGC and SGC regions, with each region further consisting of four redshift bins spanning $z=0.8$ to 2.2, each containing a similar number of sources. To analyze the inhomogeneities in the quasar distribution and explore the potential existence of homogeneity on larger length scales, we employed a multi-fractal analysis based on studying the scaling behaviour of different moments of counts-in-spheres. We first computed the correlation integral $C_2(r)$ and its logarithmic derivative, the fractal correlation dimension $D_2(r)$ across the $r$ range of (20-140) $h^{-1}$ Mpc. Additionally, to explore the higher-order clustering properties, we have also estimated the spectrum of generalized dimension, $D_q (r)$, from $C_q (r)$ for q $\in$ \{-5,-3,-2,0,3,5\}. We summarise the main conclusions from our study below,

\begin{itemize}

    \item The distribution of matter, as traced by quasars in the universe, exhibits multi-fractal behavior at length scales smaller than $80$ $h^{-1}$ Mpc.

    \item On scales exceeding ($r>80$ $h^{-1}$ Mpc), we observed the $D_2$ curve stabilizing at a constant value, typically between 2.8 and 2.9.  Our analysis confirms that both observed and mock data adhere to a random distribution within 
$1\sigma$ on these larger scales We hence conclude that the distribution of matter in the universe transitions to homogeneity beyond these scales.

      \item The variation of $D_q$ with $q$ across different length scales, establishes that the eBOSS quasar distribution exhibits homogeneity beyond $110$ $h^{-1}$ Mpc. 

      \item Also, in $D_q$ versus $q$ plots the transition to homogeneity is unambiguous at positive $q$ as compared to negative $q$. This is expected as contribution to $C_q(r)$ for negative $q$ is dominated by under-dense regions.
\end{itemize}

Thus our study where we use quasars as a tracer of underlying matter density field has demonstrated that the Cosmological Principle conjecture of a homogeneous Universe at large scales is fairly justified to a large extent. Consequently, this study is crucial in advancing our understanding of the universe's large-scale structure, formation, and evolution.

\section*{Acknowledgements}
We thank the anonymous referee for valuable comments that significantly improved the paper. PG is supported by KIAS Individual Grant (PG088101) at Korea Institute for Advanced Study (KIAS). This work is supported by the Baekdu cluster at the Center for Advanced Computation at KIAS. PG would like to acknowledge Prof Changbom Park for his useful insights and discussion. JKY and TRS thank IUCAA for the support provided through the Associateship Programme.
TRS acknowledges SERB for Project Grant No. EMR/2016/002286.

Funding for the Sloan Digital Sky Survey V has been provided by the Alfred P. Sloan Foundation, the Heising-Simons Foundation, the National Science Foundation, and the Participating Institutions. SDSS acknowledges support and resources from the Center for High-Performance Computing at the University of Utah. The SDSS web site is \href{www.sdss.org}{www.sdss.org}. SDSS is managed by the Astrophysical Research Consortium for the Participating Institutions of the SDSS Collaboration, including the Carnegie Institution for Science, Chilean National Time Allocation Committee (CNTAC) ratified researchers, the Gotham Participation Group, Harvard University, Heidelberg University, The Johns Hopkins University, L’Ecole polytechnique fédérale de Lausanne (EPFL), Leibniz-Institut für Astrophysik Potsdam (AIP), Max-Planck-Institut für Astronomie (MPIA Heidelberg), Max-Planck-Institut für Extraterrestrische Physik (MPE), Nanjing University, National Astronomical Observatories of China (NAOC), New Mexico State University, The Ohio State University, Pennsylvania State University, Smithsonian Astrophysical Observatory, Space Telescope Science Institute (STScI), the Stellar Astrophysics Participation Group, Universidad Nacional Autónoma de México, University of Arizona, University of Colorado Boulder, University of Illinois at Urbana-Champaign, University of Toronto, University of Utah, University of Virginia, and Yale University.

\section*{Data Availability}
We have utilized the publicly available SDSS eBOSS DR16 quasar catalogs for both northern (NGC) and southern (SGC) galactic caps regions provided by the SDSS-IV collaboration. They can be found at the following URL: \url{https://data.sdss.org/sas/dr16/eboss/lss/catalogs/DR16/}. Also, we have used publicly available EZmocks catalogs of the quasar sample for both NGC and SGC regions. They are given at the following URL:
\url{https://data.sdss.org/sas/dr17/eboss/lss/EZmocks/v1_0_0/realistic/eBOSS_QSO/dat/}



\bibliographystyle{mnras}
\bibliography{references} 

\bsp	
\label{lastpage}
\end{document}